\documentclass[%
 aip,
 jcp,
 amsmath,amssymb, amsfonts,%nobibnotes,
preprint,%
]{revtex4-2}
\usepackage{graphicx}% Include figure files
\usepackage{dcolumn}% Align table columns on decimal point
\usepackage{bm}% bold math
\usepackage[binary-units]{siunitx}
\usepackage{glossaries}
\usepackage[capitalise]{cleveref}
\usepackage[utf8]{inputenc}
\usepackage[T1]{fontenc}
\usepackage{mathptmx} 
\makeglossaries

\begin{document}
\newacronym{SOL}{SOL}{Scrape-Off Layer}
\newacronym{LCFS}{LCFS}{Last Closed Flux Surface}
\newacronym{ne}{$n_\mathrm{e}$}{electron density}
\newacronym{Te}{$T_\mathrm{e}$}{electron temperature}
\newacronym{IDA}{IDA}{Integrated Data Analysis}
\newacronym{MAP}{MAP}{Maximum A Posteriori estimation}
\newacronym{MCMC}{MCMC}{Markov Chain Monte Carlo}
\newacronym{MCF}{MCF}{Magnetic Confinement Fusion}
\newacronym{ITER}{ITER}{International Thermonuclear Experimental Reactor}
\newacronym{GPR}{GPR}{Gaussian Process Regression}
\newacronym{CXRS}{CXRS}{Charge Exchange Recombination Spectroscopy}
\newacronym{TS}{TS}{Thomson Scattering}
\newacronym{LOS}{LOS}{Line of Sight}
\newacronym{ECE}{ECE}{Electron Cyclotron Emission}
\newacronym{MHD}{MHD}{magnetohydroynamic}
\newacronym{FPP}{FPP}{Fusion Pilot Plants}
\newacronym{IMAS}{IMAS}{Integrated Modelling \& Analysis Suite}
\newacronym{HPC}{HPC}{High-Performance Computing}
\newacronym{TIP}{TIP}{Toroidal Interferometer-Polarimeter}
\newacronym{DIP}{DIP}{Diamagnetic Interferometer-Polarimeter}
\newacronym{NN}{NN}{Neural Network}
\newacronym{ML}{ML}{machine learning}
\newacronym{IRI}{IRI}{Integrated Research Infrastructure}
\newacronym{VVUQ}{VVUQ}{verification, validation and uncertainty quantification}
\newacronym{AUG}{AUG}{ASDEX upgrade}
\title{Simultaneous kinetic profile and magnetic equilibrium inference with Bayesian integrated data analysis in preparation for ITER}% Force line breaks with \\

\author{S. S. Denk} % Write as First name Surname
 \email[Corresponding author:]{denks@fusion.gat.com}
\author{T. B. Amara} % Write as First name Surname
\author{T. Odstr\v{c}il} % Write as First name Surname
\author{L. Stagner} % Write as First name Surname
\author{C. Akcay} % Write as First name Surname
\author{T. Slendebroek} % Write as First name Surname
\author{S. P. Smith} % Write as First name Surname
\author{M. A. Van~Zeeland} % Write as First name Surname
\author{T. Akiyama} % Write as First name Surname
\author{R. Nazikian} % Write as First name Surname
\affiliation{General Atomics, San Diego, CA, United States of America}

\date{\today} % It is always \today, today, but any date may be explicitly specified
              % Not printed for conference proceedings

\begin{abstract}
Accurate plasma state reconstruction will be crucial for the success of ITER and future fusion plants, but the harsh conditions of a burning plasma will make diagnostic operation more challenging than in current machines. Integrated data analysis (IDA) based on Bayesian inference allows for improved information gain by combining the analysis of many diagnostics into a single step using sophisticated forward models. It also provides a framework to seamlessly combine predictive modeling and data, which can be invaluable in a data-poor environment.
As a step towards integrated data analysis at scale, we present a new, fast integrated analysis framework that allows for the simultaneous reconstruction of the kinetic profiles and the magnetic equilibrium with statistically relevant uncertainties included. This analysis framework allows for the systematic evaluation of models using extensive experimental data leveraging DOE supercomputing infrastructure, such as being developed through the DOE-ASCR Integrated Research Infrastructure (Smith, XLOOP). To test the performance and verify the code it was applied to an ITER-like scenario using a realistic machine geometry and diagnostic description. Using artificial data for magnetics, Thomson scattering, interferometry, and polarimetry generated from a known ground truth, the coupled equilibrium and kinetic profile reconstruction problem was solved via the Maximum a posteriori method in approximately three minutes on a multicore CPU including uncertainty quantification. The resulting equilibrium and kinetic profiles were found to be in reasonable agreement with the ground truth. % \textbf{You must add a sentence on the resulting output and how it compares to standard efit. We don't really make this comparison at all in the paper} %\SI{3}{\minute} 
\end{abstract}

\maketitle

\section{Introduction}

Reliable and safe operation of \Gls{ITER} and future \gls{FPP} will pose significant challenges for interpretive analysis, model validation and control. One challenge is that the number of actuators for affecting the plasma state will be fewer, their coupling will be stronger, and the latency of their effects will be longer compared with existing devices. Second, extracting accurate information on the plasma state will be much more challenging than in present experiments due to the poor accessibility and harsh radiation environment of a fusion reactor compared to present experiments. These challenges will lead to difficulties in extracting accurate plasma state information with the fidelity and temporal resolution required for control and model validation. 

For example, \Gls{ITER} diagnostics, when in their pristine state, will generally yield less information and lower signal to noise than comparable systems available on existing facilities. Coupled with the anticipated degradation of various diagnostic components such as the first mirrors and lenses \cite{zvonkov2016cxrs, litnovsky2015studies} and the difficulty of regular in-situ calibration \cite{walsh2011iter,costley2005technological}, it is clear that more robust mathematical procedures will be needed for extracting maximum information on the plasma state than currently employed methods. These concerns are driving efforts to develop improved methods for plasma state reconstruction, most notably the application of Bayesian inference methods often called \gls{IDA} \cite{fischer2010integrated,verdoolaege2010potential,fischer2024integrateddataanalysisvalidation,Pavone_2023}. 

Determining the plasma state from sensor signals is a complex inverse problem due to the challenges of integrating often noisy data from heterogeneous diagnostics, the indirect relationship between the sensor signals and the plasma state parameters, and the complex coupling between state variables through the non-linear processes of the plasma. The current approach, where plasma state properties and their uncertainties are evaluated using subsets of the sensor signals and then integrated into subsequent analysis steps (the divide-and-conquer approach) has its uses but falls short when considering the need for a more systematic approach to \gls{VVUQ}, particularly in the data-poor environments anticipated for fusion reactors. 

\begin{figure}[htb]
    \centering
    \includegraphics[width=\textwidth]{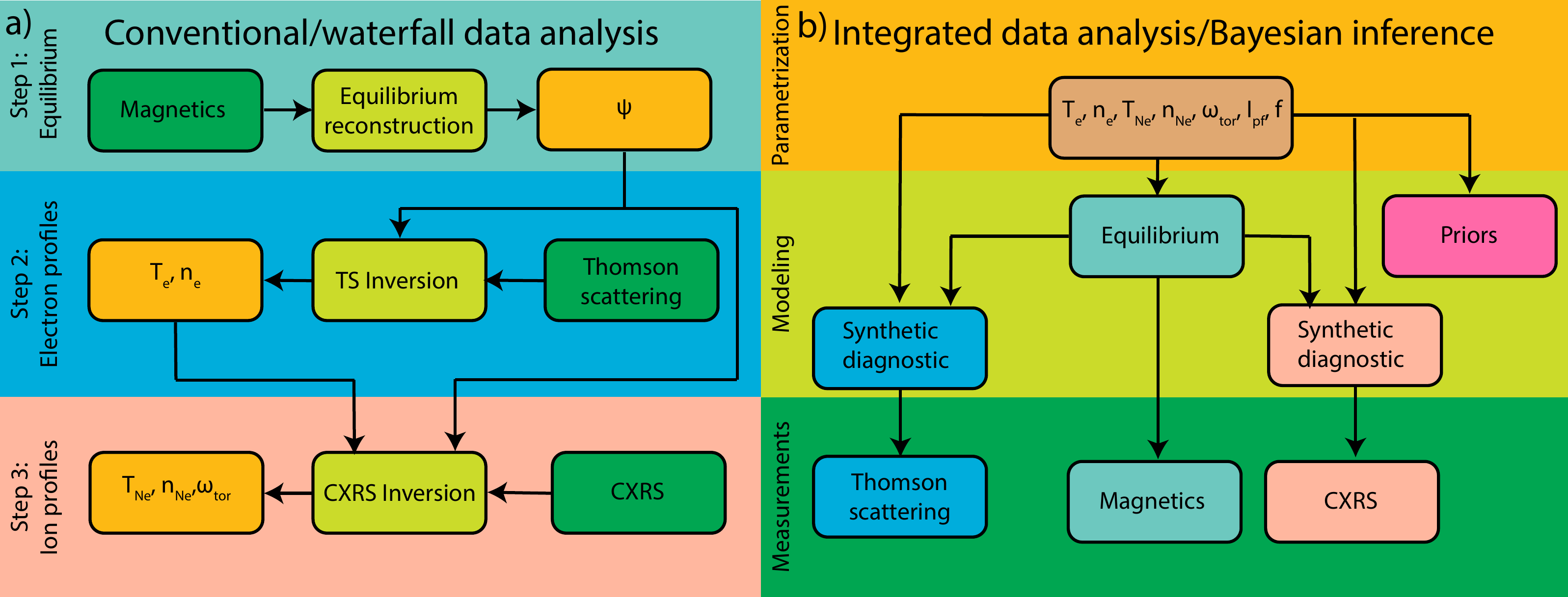}
    \caption{\label{fig:conv_vs_Bayes} Comparison between conventional or waterfall data analysis and \gls{IDA}. Figure a) shows the data flow for analyzing \gls{CXRS} and \gls{TS} measurements currently used at the DIII-D tokamak. The reconstruction and inversions are solving separate inverse problems to understand a part of the plasma state. Figure b) shows the same data flow for an \gls{IDA} approach. The equilibrium and synthetic diagnostics are forward models, and the inverse problem of plasma state determination is solved by considering an ensemble of possible plasma parameters to determine which is in best agreement with the measurements.}
\end{figure}

Here we illustrate the divide-and-conquer approach that is representative of the current practice in the field \cref{fig:conv_vs_Bayes}. First, the magnetic equilibrium is solved with a Grad–Shafranov solver (say EFIT) \cite{lao1985efit,lao2005efit}. In this step, EFIT reconstructs the equilibrium based exclusively on multiple magnetic measurements. The next step is to analyze Thomson scattering \gls{TS} data in isolation to extract electron density \gls{ne} and temperature \gls{Te} profiles using the flux coordinates extracted from the prior step for mapping the measurement locations. Importantly, EFIT is not able to predict uncertainties for the flux matrix it provides. As a consequence, the flux coordinate uncertainties do not enter into the predicted profiles from the \gls{TS} diagnostic. Another assumption in the inverse model of the \gls{TS} analysis is that the uncertainty distribution of \gls{Te} and \gls{ne} is assumed to be normal and can be characterized by an expected value and the standard deviation, which has been shown to be inadequate in the general case for fusion facilities \cite{fischer2003bayesian}. For the ions, charge exchange \gls{CXRS} measurements are analyzed in a fashion similar to the \gls{TS} measurements. However, the inference of the impurity concentration from \gls{CXRS} require \gls{ne} profiles from the \gls{TS} data. The inference of the impurity concentration does not generally include the uncertainties in \gls{TS} measurements, i.e., uncertainties are not propagated through the analysis workflow in a rigorous systematic fashion. 

A common method for addressing some deficiencies in current approach involves iteratively updating the equilibrium after evaluating the total pressure using the aforementioned procedure. While this method can yield more consistent results, it lacks rigor, fails to incorporate uncertainties in a self-consistent manner, and its outcomes vary depending on the specifics of the procedure employed by different individuals, thereby lacking overall consistency \cite{avdeeva2024accuracy}.

Although the current approach has been deemed acceptable in existing facilities where there is an abundance of high-quality data, new approaches are essential when considering the need for a more consistent and systematic approach to \gls{VVUQ}, particularly in  data-poor environments expected in fusion reactors. 

In contrast, Bayesian inference provides a mathematically consistent framework to integrate diagnostic measurements into plasma state reconstruction and uncertainty quantification \cite{fischer2003bayesian,svensson2004integrating,svensson2007large,kruger2024thinking}. This process is often referred to as Integrated Data Analysis (\gls{IDA}\cite{fischer2010integrated}). The data flow in \gls{IDA} is illustrated in \cref{fig:conv_vs_Bayes}. Bayesian inference or IDA provides a mathematical framework that incorporates prior information on the plasma state in the analysis. This information can be empirical, such as positivity for density and temperature, or estimates based on predictive models\cite{bergmann2024plasma}. All forward models (synthetic diagnostics) and the priors are combined into a single-step inverse problem that can be solved with various approaches discussed in section \ref{sect:IDA}. Each forward model can include detailed diagnostic descriptions, including calibration errors, as done for the CXRS diagnostic on \gls{AUG}\cite{nishizawa2022non}. Using this procedure, \gls{IDA} can provide a consistent and systematic approach to state reconstruction including uncertainty quantification. A more detailed discussion of the use of Bayesian inference in equilibrium reconstruction and \gls{IDA} can be found in section \ref{sect:IDA}.

On some machines, such as \gls{AUG} \cite{fischer2010integrated,IDE}, \gls{IDA} is applied routinely to every experiment to infer certain properties of the plasma state, however most diagnostic data in \gls{MCF} experiments continue to be analyzed using conventional methods. 

There are primarily two reasons for this:
\begin{enumerate}
    \item Current low-activation experiments can provide high-fidelity diagnostic data with regular in-situ calibrations, so that the need for \gls{IDA} is less urgent 
    \item \gls{IDA} of a large ensemble of diagnostic measurements is computationally intensive and more time consuming. For swift experimental decisions between discharges, it is crucial to have prompt access to inferred plasma state parameters. Delivering IDA results quickly enough to inform control room decisions has not been feasible till now. 
\end{enumerate}

In this paper we demonstrate that even a complex \gls{IDA} workflow can now be executed between experiments when combined with supercomputing resources\cite{smith2024xloop}. Leveraging leadership class computing in support of experimental facilities, such as with the emerging Integrated Research Infrastructure (IRI) developed by the DOE office of Advanced Scientific Computing Research (ASCR), and machine learning methods for surrogate models of synthetic diagnostics, IDA is tractable for plasma state determination during experimental operations. 

The paper is structured as follows. Section II provides a brief description of the \gls{IDA} framework. A description of the individual modules comprising the \gls{IDA} workflow is presented in \ref{sect:Components}. Section \ref{sect:results} presents a complete workflow for a \gls{MAP} based combined equilibrium and profile reconstruction for an \Gls{ITER} discharges. The results obtained with \gls{MAP} are then compared to \gls{MCMC} results in section \ref{sect:Verification}. In section \ref{sect:Discussion}, we discuss the potential impact of this work on VVUQ and on the operation of ITER and fusion reactors.

\section{\gls{IDA} Background}\label{sect:IDA}
Bayes' theorem provides a statistical approach to describe a relation between observations or data $D$ and model parameters $\theta$ 
\begin{equation}
 P(\theta|D) = \frac{P(D|\theta)P(\theta)}{P(D)}
\end{equation} 
$P(\theta)$ is \textit{prior} information about the parameters, such as smoothness, positivity, or monotonicity.  $P(D|\theta)$ is the \textit{likelihood}, i.e. the probability of observing the data $D$ given the parameters $\theta$, and  $P(D)$ is the marginalized evidence $P(D) = \int_\theta P(D|\theta)P(\theta)$, which serves as a normalization constant in this context. The likelihood $P(D|\theta)$ establishes a connection between the parameters and measured data based on forward models and synthetic diagnostics. Finally, the posterior $ P(\theta|D)$ of the model parameters $\theta$ describes the degree of belief we have in
the parameters, which is based on prior knowledge and updated by the observed data. The posterior encapsulates all of the information and uncertainties on the model parameters.

Calculating the posterior probability for an initial parameter set is only the first step. Although one could evaluate the posterior on the parameter grid, this task becomes nearly impossible in high-dimensional space. For a problem with 100 parameters, a regular sampling on a 100-dimensional hypercube with just ten samples per dimension requires $10^{100}$ (googol) samples. Two different methods for exploring the posterior distribution are commonly employed - \gls{MCMC} or \gls{MAP} estimate. \gls{MCMC} is a class of algorithms used to draw random samples from a posterior distribution. These algorithms are particularly suited for high-dimensional problems since they suffer less from the curse of dimensionality. However, \gls{MCMC} algorithms still require a significant number of posterior evaluations, and many \gls{MCMC} algorithms cannot be efficiently parallelized. For routine data analysis, \gls{MAP} is more suitable as it is orders of magnitude faster and finds the most probable parameter ensemble for the given measurements. 

The downside of \gls{MAP} is that its uncertainties are derived from the local curvature of the posterior by approximating the posterior as a multivariate Gaussian. This poorly describes the uncertainty of skewed, multi-modal, or otherwise non-Gaussian distributions where the mean and median may not align \cite{rasmussen2003gaussian}, but \gls{MAP} can still be used for problems where these distributions are possible if extra care is taken. For skewed distributions, as the number of data samples is increased, the uncertainty is reduced, and a local approximation becomes more accurate.
With multi-modal problems, the best option is to use additional data sources that select one of the peaks or otherwise bring the distribution closer to Gaussian. The \gls{MAP} approach has proven effective for non-linear problems in data-rich environments \cite{fischer2003bayesian}, but it is not guaranteed. Additionally, it is often possible to transform the distribution to be more Gaussian with proper variable transformations. This requires a good understanding of the possible distributions for a problem which can only be confirmed with MCMC.

Feasibility of such inference was demonstrated for a case including electron kinetics profiles only and magnetics equilibrium on JET \cite{kwak2022bayesian}. In that work, the kinetic profiles and magnetic equilibrium were described by Gaussian Processes, and the Grad-Shafranov equation played the role of a virtual observation included as a part of prior knowledge. Although such an approach is feasible, it is not practical due to an enormous computational cost, which will increase even further with additional diagnostics. Like this approach, we will not directly solve the Grad-Shfarnov equation with a forward model but use a surrogate that approximates the solutions.  

Rather than use a model based on a simplified physical description, we employ \gls{ML} surrogate models to replace expensive computations.
This is similar to the approach developed on W7-X, where the magnetics equilibrium was calculated by an artificial neural network (ANN) \cite{merlo2023accelerated}. The approach allowed for fast and robust inference of electron profiles and the 3D \gls{MHD} equilibrium of the stellarator.
In this work, we have expanded those ideas to use a much more general framework that also infers the plasma current density by fitting magnetic measurements and coil currents and allows for the inference of ion profiles. In addition, we use generic B-splines to represent the profiles rather than assuming a particular shape, and we use a gradient-based optimization for MAP that is enabled by the vectorized evaluation of many different parameters. This allows for a faster, generalized, and more robust \gls{IDA} procedure as well as closer connections with the ITPA initiatives \cite{gonzalez2023itpaida,fischer2024integrateddataanalysisvalidation}.

%In this work we will demonstrate the ability of \gls{IDA} to infer the plasma state directly from the calibrated raw data without any intermediate steps. %TBA: possibly a bit too repetitive
%However, such inference is difficult due to the size and complexity of this task \cite{nishizawa2022impurity}.
Directly inferring the complete plasma state with \gls{IDA} is difficult because of the size and complexity of this task.% \cite{nishizawa2022impurity}. %TBA: this is a more streamlined intro
The number of routine measurements available on tokamaks is relatively large; particularly when raw data from spectral diagnostics are included. For example, the DIII-D \gls{CXRS} diagnostic provides $\sim 10^5$ raw measurements, one for each pixel of a CCD camera. The core plasma state is described by profiles of the density, temperature, and poloidal and toroidal rotation for all plasma species, including fast ions and neutrals, the plasma current, poloidal flux matrix, etc. 
Even if the temperature and rotation are assumed to be equal for all thermal ions, at least ten radial profiles are still required. A non-parametric representation of the profiles by Gaussian processes capable of capturing steep pedestal and internal transport barrier gradients requires $\approx10^2$ points \cite{nishizawa2022impurity}, so fitting all of the profiles together would require $\approx10^3$ dimensions. Often systems developed for solving the inverse problem use reduced resolution, e.g., real-time EFIT initially solved for the poloidal flux on a 33x33 grid. Our goal is to use sufficient resolution to capture millimeter deviations in the flux surfaces on \Gls{ITER} which requires a grid on the order of 300x300, bringing the total dimensionality of the plasma state to $\approx10^4$.
Determining the posterior of such a high-dimensional Gaussian process would be feasible for a linear problem, but the inverse problem in IDA is non-linear.
Instead of reducing the spatial resolution, we rely on higher order or spectral parametrizations so that we can reduce the problem dimensions without sacrificing accuracy.

\section{Components of the \gls{IDA} framework}\label{sect:Components}
%TODO: This needs to be improved. Is it necessary to go in the implementation details? 
%should we even start by this? 
The main result presented in this paper is a new, modular \gls{IDA} framework that is discussed in detail in this section. It is constructed as a collection of Python language-based components whose coupling is specified through a configuration file defining a data analysis workflow. It uses the \Gls{ITER} \Gls{IMAS} library to load and save data, and the \gls{IMAS} schema is used internally to exchange data between components. Presently, there are eight categories of components:
\begin{enumerate}
    \item The \underline{core} of the \gls{IDA} framework runs \gls{MAP} and \gls{MCMC} and propagates the uncertainties of parameters;
    \item The \underline{posterior} and \underline{posterior dispatcher} assembles the components and calculates the posterior function, which can be distributed across multiple processors;
    \item \underline{Parametrizations} translate between low- and high-dimensional representations of plasma parameters that are either directly inferred by \gls{IDA} or derived by one of the codes;
    \item \underline{Generic code components} that either compute synthetic diagnostics or derive additional quantities needed by other components;
    \item \underline{Diagnostics} manage the loading of measurements and their mapping to synthetic diagnostic signals;
    \item \underline{Likelihoods} combine the measurements with synthetic data computed by code objects;
    \item \underline{Priors} constrain the inferred parameters;
    \item The \underline{IDSServer} reads, stores, and writes data using the \gls{IMAS} library and facilitates the data exchange between components;
\end{enumerate}
Each of these is described in detail in the subsections below.

\begin{figure}[h]
    \centering
    \includegraphics[width=\textwidth]{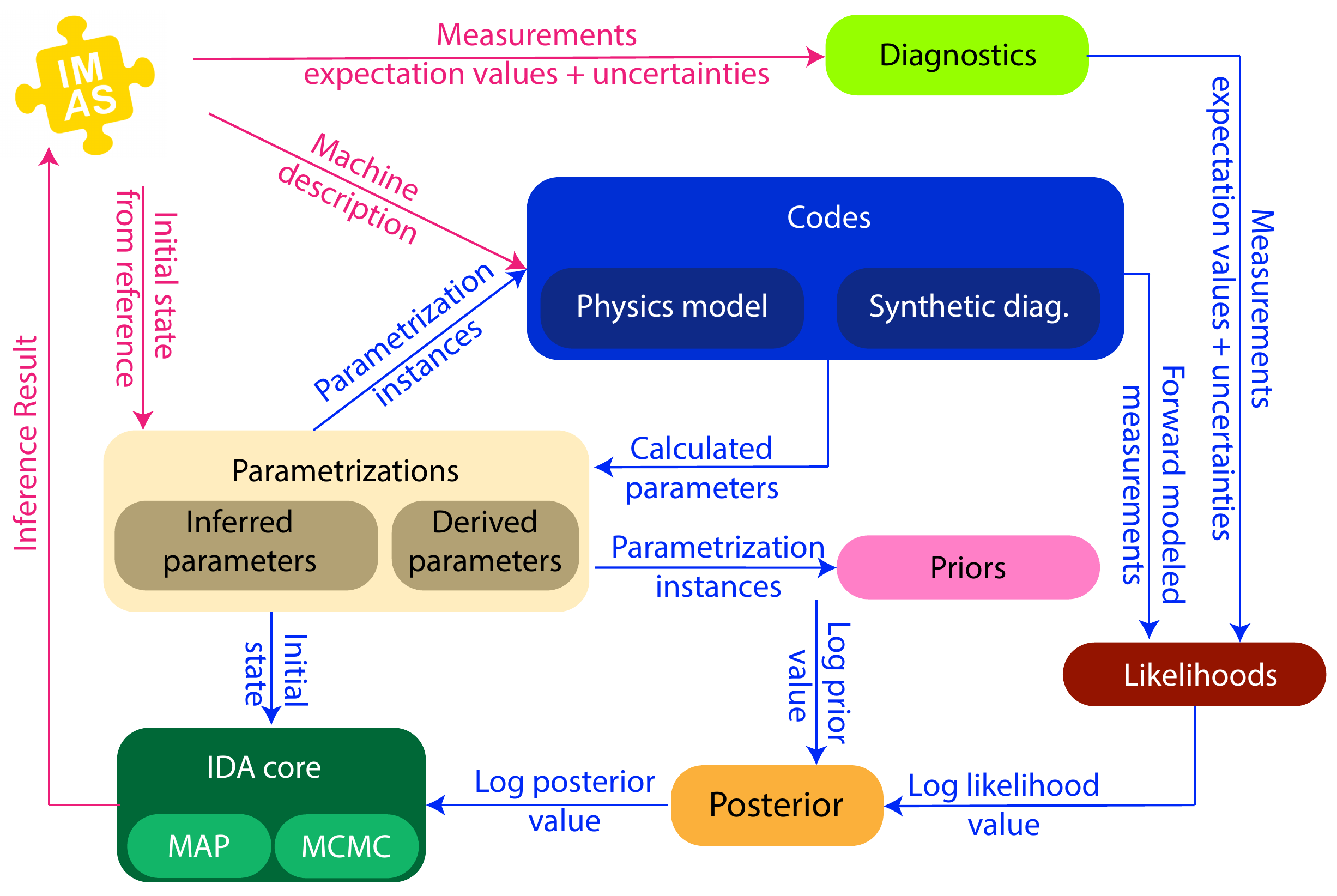}
    \caption{\label{fig:IDA_data_flow} This figure illustrates the data flow between the various components in the \gls{IDA} framework. The individual components load diagnostic data, the machine description, and an initial guess for the inferred parameters from an \gls{IMAS} database. The optimizer propagates each guess through the code via \textit{parametrization} components. \textit{Codes} provide synthetic data for likelihoods and derive further parameters. The \textit{likelihoods} combine the synthetic measurements from \textit{codes} with the measurement data from \gls{IMAS} in a likelihood object. Finally, each \textit{prior} is added to the posterior.}
\end{figure}

\Cref{fig:IDA_data_flow} illustrates the data flow in the \gls{IDA} framework.
% The user provides an initial guess in the form of a reference scenario. Each parameterization object loads parameters from the scenario as part of its initialization. Code objects load the machine descriptions they need, calculate synthetic diagnostic data and derive additional parametrizations that other codes or priors need. Diagnostic objects load measurement data and compute the likelihood from the synthetic measurements. The updated posterior, the sum of log-likelihood and log-prior, and its Jacobian (not depicted) are then used by the \gls{IDA} core optimization to provide a new set of sample parameters to be tested. In the following subsections, we will describe each component of the \gls{IDA} framework in more detail.

\subsection{The core of the \gls{IDA} framework}
The primary purpose of the \gls{IDA} core is to perform the inference via \gls{MAP} or \gls{MCMC} and to propagate the uncertainties of the inferred parameters to the actual parameters of interest (e.g. from B-Spline coefficients to asymmetric error bars of profiles). %It also reads in the configuration file and instantiates the posterior dispatcher class.

\subsubsection{MAP optimization of the posterior}
The \gls{MAP} problem is solved via the BFGS (Broyden, Fletcher, Goldfarb and Shanno) method, a gradient-based second-order optimization scheme that estimates the Hessian iteratively from the log-posterior gradient\cite{fletcher2000practical}. The gradient of the posterior is calculated with numerical differentiation employing vectorization and parallelization across the various samples. It allows switching the numerical derivatives between the forward, centered, or five-point stencil schemes, requiring $N_\mathrm{param} + 1$, $2\times N_\mathrm{param}$, or $4\times N_\mathrm{param}$ posterior evaluations, respectively. For all results shown in this paper, the five-point stencil is used to ensure numerical accuracy. In addition, the final sum of all log-posterior contributions is evaluated in quadruple precision (128-bit floating point) to minimize the discretization error. Gradient evaluations can be significantly accelerated by assuring the posterior functions and all its constituents are fully vectorized and parallelized with respect to the parameter ensemble. For details on the parallelization, see section \ref{sect:posterior_dispatcher}.

\subsubsection{\gls{MCMC} analysis}
While too slow for routine analysis, an \gls{MCMC} approach is invaluable for verifying the \gls{MAP} results. Two \gls{MCMC} packages have been setup in the \gls{IDA} framework: the ensemble step sampler EMCEE \cite{EMCEE} and the ensemble slice sampler Zeus \cite{karamanis2021zeus,karamanis2020ensemble}. This work presents only results from EMCEE since it is more parallelizable than Zeus and similarly efficient when drawing samples. The \gls{MCMC} chains are initialized from the \gls{MAP} solution, which significantly accelerates the convergence, particularly for high dimensional problems.

While \gls{MAP} usually requires less than $10^4$ posterior evaluations, getting a converged \gls{MCMC} chain requires several hundred thousand steps with several hundred to a thousand walkers, leading to $10^7-10^8$ posterior evaluations. While the computational efficiency of the \gls{IDA} framework is important so the \gls{MAP} approach can deliver results fast enough for in-between-discharge analysis, the computational efficiency, parallelization, and vectorization are crucial for the \gls{MCMC} approach to compute converged chains at all.

\subsubsection{Uncertainty propagation}
\label{sect:uncertainty_propagation}
For a \gls{MAP} analysis, the uncertainties are propagated by drawing samples from a multivariate Gaussian using the inverse Hessian posterior matrix as its covariance. This allows uncertainty propagation to all quantities calculated by the \gls{IDA} framework. \gls{MCMC}, on the other hand, provides the samples directly. The uncertainties are visualized in the result figures below as \num{15} and \num{85} percentiles, plotted as asymmetric error bands. For the expected value, the mode is used in the case of \gls{MAP} results and the median for \gls{MCMC} results.

\subsection{The posterior and posterior dispatcher component}
\label{sect:posterior_dispatcher}
The next component is the posterior and the posterior dispatcher. The dispatcher uses the configuration file to assemble the pipeline required to evaluate the posterior. The posterior dispatcher uses the Python \verb|multiprocessing| package to parallelize posterior evaluations across multiple CPUs. While many calculations in the \gls{IDA} framework are performed by libraries like \verb|numpy| and \verb|tensorflow| that feature parallelization internally, there is still performance to be gained when evaluating large batches of samples in a single step. The core of the \gls{IDA} framework passes a large batch, e.g. \num{1000} samples for \gls{MCMC} with \num{1000} walkers, to the dispatcher that then splits it into smaller batches more suitable for vectorization across a single CPU and to reduce the number of cache-miss events. Each process has its own instance of the posterior and all its respective required components. Data is exchanged through the individual instances of the IDSServer (described in section \ref{sect:IDSServer}) and --- if more than just the posterior values are needed (for example, for the uncertainty propagation) --- the dispatcher can merge the results from the various parallel instances into a single IDSServer instance.

\subsection{Parametrization components}\label{sect:parametrisation}
The next component of the \gls{IDA} framework are the parametrizations.
The general purpose of parametrization components is to map the low-dimensional state vector to a more useful, high-dimensional representation. For the inference it is key to minimize the dimensionality of the problem for computational performance and to reduce overfitting. However, for the physics model, such as synthetic diagnostics, it is important to be able to evaluate the physics parameters they need. In addition to the expansion of low-dimensional representations, the parametrizations also handle the storing and loading of the low-dimensional parameters through the IDSServer (see section \ref{sect:IDSServer}) and the evaluation on fixed grids for error propagation and visualization.

\begin{description}
\item[Direct parametrization] 
The simplest form of parametrization has its parameters directly inferred by the optimizer. Examples would be scalar plasma parameters like the vacuum magnetic field, or vectors like the poloidal field currents $I_\mathrm{PF}$, which are an input for the equilibrium surrogate model E-Forward-NN (see section \ref{sect:EForward-NN}).

\item[Profile parametrization] 
One-dimensional profiles like $T_\mathrm{e}$ or $n_\mathrm{e}$ are parametrized by 3rd order B-splines evaluated at normalized poloidal flux $\Psi_\mathrm{N}$. For positive quantities, such as temperatures or densities, the spline $s(x)$ is transformed using the $\exp()$ function to enforce positivity and reduce the magnitude spanned by the spline coefficients. This approach was proven to be successful by ref. \citenum{fischer2010integrated}. The derivative is evaluated with the chain rule to reduce numerical errors (e.g. $\exp\left[s(x)\right]' = \exp\left[s(x)\right] s'(x)$).
% The advantage of $\Psi_\mathrm{N}$ compared to other commonly used radial coordinates like $\rho_{\Psi}$ and $r/a$ is that $\Psi_\mathrm{N}$ is properly defined outside of the separatrix and pedestal gradients are less steep compared to $\rho_{\Psi}$. % and profiles do not have zero derivatives on axis. - TBA: this is probably a disadvantage...
B-spline knot locations are fixed, and their locations were estimated from a training database to provide the best representation of the generated profiles with the lowest number of knots. Pressure profiles, $p'$, and $ff'$ splines use 12 knots and have individual locations. All pressures share the basis with $p'$. In a similar fashion a separate set of 12 spline knot locations optimal for \gls{Te} and \gls{ne} was obtained. For all quantities but $ff'$, a high knot density near the edge is crucial to capture the steep pedestal region adequately.
%\cref{fig:TeBasis} illustrates the optimal B-spline basis for $T_e$ profile. The higher knot density in the edge is essential for accurate description of the steep pedestal region.

The vectorized evaluation of multiple sets of spline coefficients is accomplished by multiplying the spline coefficients with the basis, precalculated on a fixed grid, followed by a linear interpolation at diagnostics' $\Psi_\mathrm{N}$ locations. This method also allows the vectorized evaluation of derivatives and integrals using precalculated basis function gradients and antiderivatives of B-splines.

% \begin{figure}
%  \centering
%  \includegraphics{fig/spline_basis_Te.pdf}
%  % spline_basis_Te.pdf: 0x0 px, 0dpi, nanxnan cm, bb=
%  \caption{Example of the B-spline basis for electron temperature profile. }
%  \label{fig:TeBasis}
% \end{figure}

% For the uncertainty propagation, the B-spline is evaluated for all provided parameter samples, and the upper and lower error bars of the profiles given by the \SI{15}{\percent} and \SI{85}{\percent} percentiles of the evaluated profiles.

\item[Parametrization of the Poloidal Flux Matrix] 
The poloidal flux matrix is parametrized by a $16\times34$ set of the Chebyshev polynomials. The Chebyshev polynomials allow for reasonably fast vectorized evaluation of multiple flux matrices in one step and analytical calculation of the gradients to derive a local magnetic field. While \verb|numpy.polynomial.chebyshev| allows multiple sets of coefficients to be evaluated in a single call, this process is still relatively slow. Hence, Chebyshev basis functions are evaluated at predefined $R,Z$ coordinates of diagnostics measurements, and polynomial evaluation is turned into a substantially faster matrix-matrix multiplication. This approach is particularly efficient because, unlike the $\Psi_\mathrm{N}$ positions of diagnostics, the $R$ and $Z$ locations do not change during the optimization. Despite this, evaluating this matrix multiplication is still expensive because of its high dimensionality. For example, updating the $\Psi_\mathrm{N}$ coordinates on the \gls{LOS} of the interferometry requires multiplication of $N_\mathrm{samples}\times544$ and $544\times2300$ element matrices, where $N_\mathrm{Samples}$ can be as large as \num{200}.
\end{description}

\subsubsection{Parameter transformations}
\label{sect:transformations}
A dedicated transformations component translates the parameters from near unity optimizer space to physics space, with units according to the \gls{IMAS} schema. This transformation is necessary to keep the guesses from \gls{MAP} and \gls{MCMC} inside the training region of the \gls{NN}s (see section \ref{sect:EForward-NN}). Each parameter is transformed individually from unbounded optimizer to bounded physics space via the cumulative normal distribution function multiplied by a parameter range that covers 98\,\% of the cases in the training database. All parameters are treated as uncorrelated, so each range is determined on a per-parameter basis (e.g., each spline coefficient separately), and physics space has thus the shape of a hypercube. Since the training data itself is heavily correlated, this does not prevent out-of-training parameter guesses. However, it reduces them so much that catastrophic failures of the \gls{IDA} framework are rare. This transformation is accompanied by an appropriate prior to prevent the optimization from running to infinity in optimization space (see section \ref{sect:priors}).

\subsection{Codes}\label{sect:codes}
Any component that uses parametrizations to calculate either synthetic measurements or derive further parametrizations is classified as a code component in the \gls{IDA} framework. A good example is our model for the equilibrium reconstruction, which functions as a synthetic diagnostic for magnetic measurements but also delivers many other equilibrium quantities, such as the parametrization of the flux matrix.

\subsubsection{MUSCLE3 adapter}

The framework fully supports coupling external codes through MUSCLE3 \cite{muscle3} using version 0.7.1. With this adapter the \gls{IDA} framework is in principle ready to couple to any \gls{IMAS} actor that supports MUSCLE3. A good future use-case would be coupling to an \gls{ECE} actor like ECRad \cite{ECRad} or SPECE \cite{SPECE}, both of which already have a MUSCLE3 interface. At present, this adapter is not used directly in the current analysis workflows, but it was applied to couple to the boundary fitting method in EFIT \cite{lao1985efit,lao2005efit} when generating the database for the E-Forward-NN surrogate model development. MUSCLE3 allows us to quickly change any forward model e.g. we can switch between predictive EFIT and the E-Forward-NN surrogate.

\subsubsection{Profile and map models}
The profile and map models are simple codes that have the purpose of reading either $\Psi_\mathrm{N}$ or $R,Z$ coordinates from a source \gls{IMAS} field specified in the configuration, evaluating a predefined parametrization object, and then storing the results in another \gls{IMAS} field. There are three varieties in total: a generic model for 1D profiles, a model for 2D parametrized flux matrices, and a third model that derives the magnetic field from parametrized flux matrices.

An example for the map and profile models is the trivial synthetic \gls{TS} diagnostic. The $R,Z$ coordinates of each channel are loaded from a machine description file, the map model then computes the $\Psi_\mathrm{N}$ positions. The profile model 
then interpolated the parametrization for \gls{Te} and \gls{ne} at the measurement locations of the \gls{TS} system.

\subsubsection{Pressure model}
This model calculates the pressure for each considered species and adds any pressure that is parametrized directly. It has two modes; it either sums up the pressure across all species to produce the total pressure or if a spline parametrizes the total pressure, it subtracts the pressure of each parametrized species from the total pressure to compute a residual pressure. The latter mode is applied for the example in section \ref{sect:results}, where the ion pressure is derived from the total pressure and electron pressure.

\subsubsection{Profile integration model}
Since the equilibrium forward model described in the following section depends on $p'$ and $ff'$, but $p$ and $f$ profiles are required for the pressure model and calculation of the toroidal magnetic field, a model that performs the integration is required. The quantities are integrated using pre-computed basis functions of the B-spline antiderivatives.  The integration constants can be either a free parameter of the inference or an external reference value, like $B_\mathrm{vac} \cdot R_\mathrm{B,ref.}$ for $f$.

\subsubsection{EForward-NN}\label{sect:EForward-NN}
EFIT \cite{lao1985efit,lao2005efit} is the most popular code for reconstructing tokamak equilibria and is used on the majority of tokamaks worldwide. It computes free boundary Grad-Shafranov (GS) fits by minimizing chi-squares, assuming all diagnostics/constraints have normally distributed errors or are fixed. As such, it solves the same inverse problem that is presented in section \ref{sect:mag_recon} with a different algorithm. It does not, however, produce uncertainties for any fit variables or computed parameters. We aim to replace this process with the Bayesian methods in \gls{IDA}, but to do so, we need a forward model for the equilibria given a minimal plasma and external coil description. However, this option is not available in EFIT or any iterative Grad-Shafranov solver because the vertical instability of elongated plasmas requires some feedback mechanism to counteract \cite{jardin2010methods}. The most similar option in EFIT is the boundary fitting (predictive) method, which varies the coil currents to find a GS solution that is the closest possible match to a chosen plasma shape, with the $P'$ and $FF'$ profiles specified.

This boundary fitting option is inconvenient in \gls{IDA} for several reasons. For one, there are no low-dimensional parametrizations of plasma shapes, including the X-point and open magnetic surfaces that are general enough to describe all possible tokamaks. For example, the commonly used Miller-extended harmonic (MXH) representation\cite{arbon2020mxh}, requires a high number of Fourier coefficients to capture an X-point. Moreover, the MXH parameters must be carefully constrained to avoid problematic shapes, such as self-intersections. The second issue is that GS solutions may only exist for some combinations of boundaries and plasma profiles, so the parameter space needs to be further restricted. Knowing how to define the feasible parameter space while also allowing for all possible tokamak configurations is non-trivial. A third challenge is that EFIT is too slow for inference ($\approx1$ second) even when running this way. Lastly, since the solutions are calculated iteratively, they are not differentiable such that performing a gradient-based search over parameter space is not feasible for \gls{IDA}.

We have developed a \gls{ML} surrogate EFIT forward model neural network (E-Forward-NN) using techniques developed in the EFIT-AI project \cite{mcclenaghan2024efitai} to address these issues. Like in that work, the E-Forward-NN only predicts the flux produced by the plasma ($\Psi_{plasma}$) without any contribution from the coils. The important difference is that the current model was trained to predict the coefficients of the Chebyshev expansion of $\Psi_{plasma}$, rather than PCA components of the current. This choice was motivated by consistency with the aforementioned parametrization but could be replaced in the future.
Similarly, the model inputs include the B-spline coefficients for $P'$ and $FF'$, where the prime denotes derivatives with respect to the normalized poloidal flux ($\Psi_\mathrm{N}$), which differs from the EFIT inputs that use the derivatives with respect to $\Psi$ (note: the EFIT code was updated to use B-splines to ensure consistency for this project). The other significant difference between E-Forward-NN and EFIT is that rather than explicitly using a target boundary shape as the input, E-Forward-NN uses a set of external coil currents, and separatrix shape is defined implicitly by $\Psi$. These input parameters are more straightforward to vary to capture the parameter space of possible solutions and they are general enough to include every reasonable configuration. The coil currents are treated as a measured quantity in the \gls{IDA} framework, and like with all measurements, a finite uncertainty is allowed. This is particularly important for the equilibrium reconstructions since eddy currents in the vessel aren't directly being modeled by EFIT but can generally be captured by variations in the shaping coil currents \cite{lao1985efit}.

Rather than constructing E-Forward-NN as a single densely connected \gls{NN}, we have trained several different \gls{NN}s that all use the same inputs but infer parameters with different scales. Each model is a densely connected multi-layer-perceptron with two hidden layers using Swish activation functions (SiLU). Separate models are used for producing 1) the Chebyshev coefficients of $\psi_{plasma}$, 2) positions of the magnetic axis and x-points, 3) synthetic magnetic sensor and toroidal flux loop measurements, and 4) total toroidal plasma current ($I_p$). $\psi_{plasma}$ is added to the external coil fluxes ($\psi_{ext})$ to produce the total flux ($\psi = \psi_{plasma}+\psi_{ext}$), which is an essential output for mapping other diagnostics such as \gls{TS}. With the magnetic axis and x-point positions, the flux can be normalized without expensive fieldline following calculations. The \gls{IDA} framework also has the option to compute synthetic measurements and $I_p$ directly from the plasma current distribution ($j$), using Green's functions to store the diagnostic responses, but this is significantly slower to compute with the parametrization being used.  Alternative parameterizations that could make these computations faster will be considered in future work.

\begin{table}[]
    \centering
    \begin{tabular}{|c|c|}
        \hline
        Inputs (for every NN)\\
        \hline\hline
        $P'$ \\
        \hline
        $FF'$ \\
        \hline
        External coil currents \\
        \hline
    \end{tabular}
    \hspace{1cm}\begin{tabular}{|c|}
        \hline
        Outputs (each from a different NN)\\
        \hline\hline
        Chebyshev coefficients of $\psi_{plasma}$ \\
        \hline
        Magnetic axis and active x-point positions\\
        \hline
        Synthetic magnetic sensor and\\ toroidal flux loop measurements \\
        \hline
        Total toroidal plasma current ($I_p$) \\
        \hline
    \end{tabular}
    \caption{\label{tab:NN_IO} This table lists the inputs and outputs of the different NNs that comprise E-Forward-NN.}
\end{table}

E-Forward-NN was trained from a database of about \num{50000} possible \Gls{ITER} equilibria generated for this project via the FUSE integrated design tool \cite{meneghini2024fuse}. Each equilibrium represents a different transport steady state based on the heating and current drive systems using the machine description files from the \Gls{ITER} \gls{IMAS} database at full field (5.3T). For a detailed breakdown of the considered machine description files please see \cref{tab:diag_MD}. The target plasma shapes used to generate these cases in FUSE are based on all DIII-D experiments conducted in the past 5 years after scaling to the \Gls{ITER} shape. Possible plasma shapes were sampled using an invertible \gls{NN} and down selected for cases with a lower single null inside the \Gls{ITER} wall. The profiles in each FUSE solution were then randomly perturbed 10 times to create a dataset of about \num{500000} cases that better represents the possible search space for \gls{IDA}. Finally, these cases were all run with EFIT, using the MUSCLE3 adapter described above, to generate magnetic sensor data. Ultimately, this yielded \num{376166} potential training cases. % TBA: I don't think it's worth going into this much detail about why cases were lost (particularly since there isn't any reason to expect all the wiggled cases to converge), but if you do then the following should be mentioned: 1. FUSE used a fixed boundary equilibrium with a separatrix added later, 2. the convergence criteria chosen for EFIT were as stringent as possible without losing too many cases, and 3. spline fitting of FUSE data could be affecting the translation to EFIT

\begin{figure}[h!]
    \includegraphics[width = 0.8\textwidth]{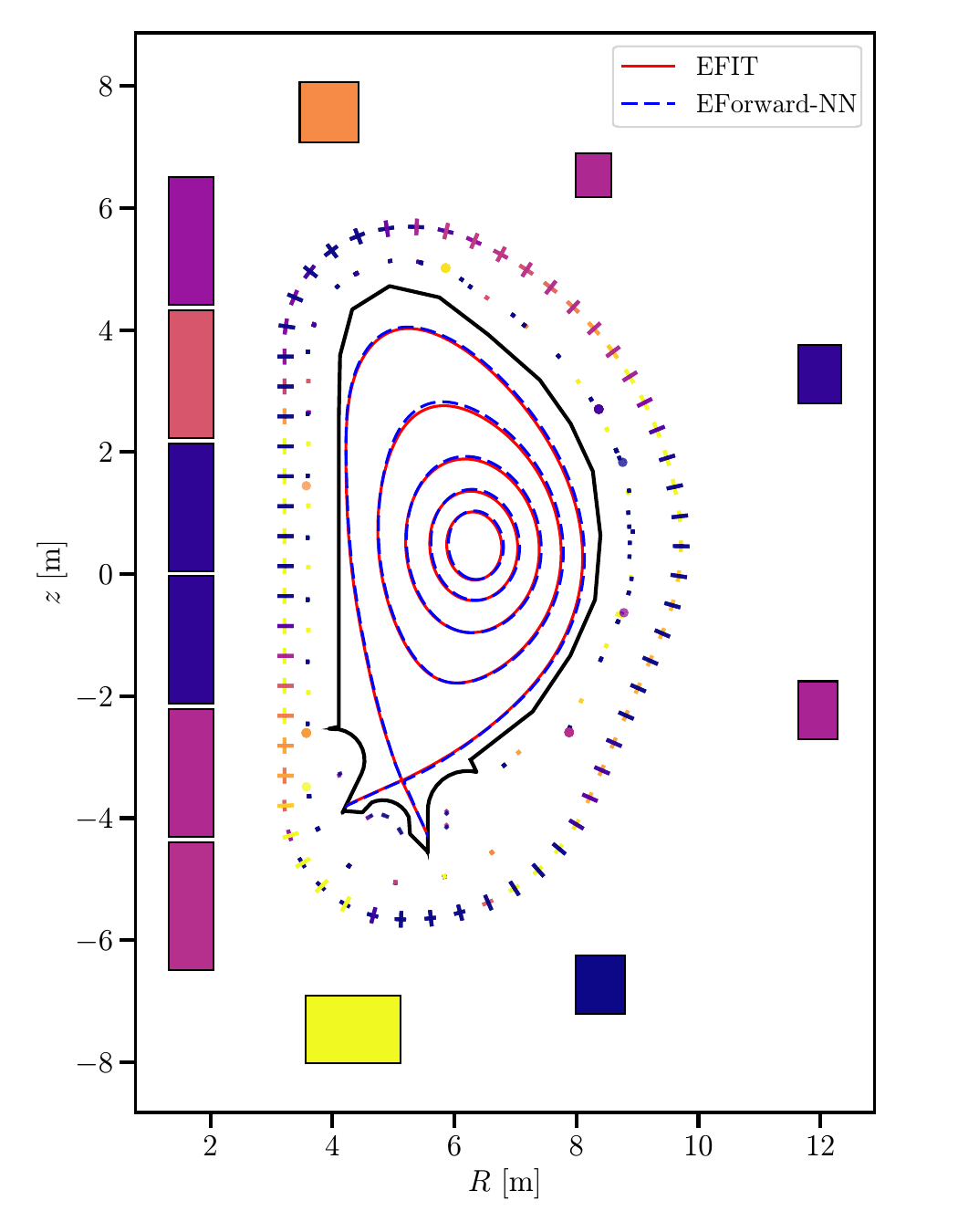}
    \caption{\label{fig:NN_comparison} The flux surfaces inferred by E-Forward-NN are in reasonable agreement with EFIT for the case of the final result of the kinetic reconstruction presented in section \ref{sect:kinetic_reconstruction}. The same holds for magnetic diagnostics, whose magnitude is shown by the colored symbols outside the wall. The rectangles with black outlines are the external coils (both shaping and the center solenoid), the circles are flux loops, and the lines are the magnetic pick-up coils (which appear as crosses because of overlapping probes with different directions).}
\end{figure}

Each NN model was trained on this data using the Tensorflow framework with a 70:15:15\% train:validate:test split until converged. A comparison of the E-Forward-NN inference to EFIT for the target case presented in section \ref{sect:results} is shown in \cref{fig:NN_comparison}.
The agreement is reasonable for mapping the diagnostics in the inference. Compared to the original EFIT calculation time, E-Forward-NN is about \num{1000} times faster, especially when the inference is batched, and many samples are evaluated at once. 

While it would be more convenient and efficient to forgo the integration model, and train E-Forward-NN using the total pressure $p$ and the diamagnetic function $f$, the performance of the resulting \gls{NN} is worse than one equivalently trained on $p'$ and $ff'$. In order to avoid sampling outside the trained region of E-Forward-NN, $p'$ and $ff'$ are subject to the parameter transformations described in section \ref{sect:transformations}. 

\subsubsection{Interferometry and Polarimetry synthetic diagnostics}
The Interferometry and Polarimetry synthetic diagnostics use formulas (1-5) from ref. \citenum{van2017tests}, retaining the second-order electron temperature corrections for the vibration compensated phase $\phi_\mathrm{intf}$ and polarimeter phase shift $\phi_\mathrm{pol}$. The integration along the \gls{LOS} is performed with Simpson's method using a uniform step size of \SI{5}{\centi\meter}.

\subsection{Diagnostics}
Because the synthetic diagnostics are implemented in the codes described above, the only remaining step is the loading and preparation of data and selection of the likelihood in the diagnostic component.
This is trivial for the examples considered in this work because purely artificial data in \gls{IMAS} format are used. 
In the current workflow, a configuration fully specifies three required items: The \gls{IMAS} fields to load measurement data and uncertainty from, the \gls{IMAS} field where the corresponding synthetic diagnostic code stores its prediction, and the type of likelihood that will be used. So, for example, the diagnostic would load \verb|thomson_scattering.channel.t_e.data| and \verb|thomson_scattering.channel.t_e.data_uncertainty_upper| and then connect these two values to \verb|thomson_scattering.channel.t_e.reconstructed| --- which does not exist in IMAS 3.41.0 --- in the likelhood. % TBA: it might be worth giving an example here. It probably isn't obvious to anyone who isn't an IMAS expert that the measured and synthetic diagnostics should be stored in different IMAS entries.
%Therefore, anything diagnostic-specific is encapsulated into the synthetic diagnostic code described in the previous section, and every diagnostic considered in this work uses the same generic implementation for the diagnostic component itself. % TBA: this is probably too far in the implementation weeds and is confusing without being able to look through the code.
In order for the \gls{IDA} framework to be applied to experimental data, this class will need to be extended to include options to slice, filter, or apply other required transformations.
This pre-processing is not necessary with the artificial data considered here and is left for future work.

\subsection{Likelihoods}
Currently, only Gaussian and Cauchy likelihood components are implemented. Like any other component of the framework, the likelihoods are fully vectorized and capable of evaluating multiple parameter samples at once. While the Cauchy likelihood is implemented it is not used for the results shown in section \ref{sect:results} because all artificial signals are generated with Gaussian noise.

\subsection{Priors}\label{sect:priors}
Priors represent beliefs about the inferred parameters. They are primarily used to restrict the posterior to more sensible physics solutions. All priors are currently empirical, but utilizing physics-based priors is planned \cite{bergmann2024plasma}. For now, the priors only act on the B-spline interpolated on a fixed grid. The three different types of priors currently implemented --- curvature, positivity and monotonicity, and parameter boundary constraints --- are described below.

One of the biggest challenges with empirical priors is that they introduce tuning hyperparameters that for now need to be adjusted manually.
At the very least, their magnitude needs to be roughly on the same order as the likelihood from the measurements.
This particular scaling has been automated by multiplying the priors with the number of measurements.
An exception to this scaling is the boundary prior, which scales with the number of parameters instead.
%Say something about planned hyperparameter optimization? 

\subsubsection{Curvature constraint}
\begin{figure}
    \includegraphics[width = 0.5\textwidth]{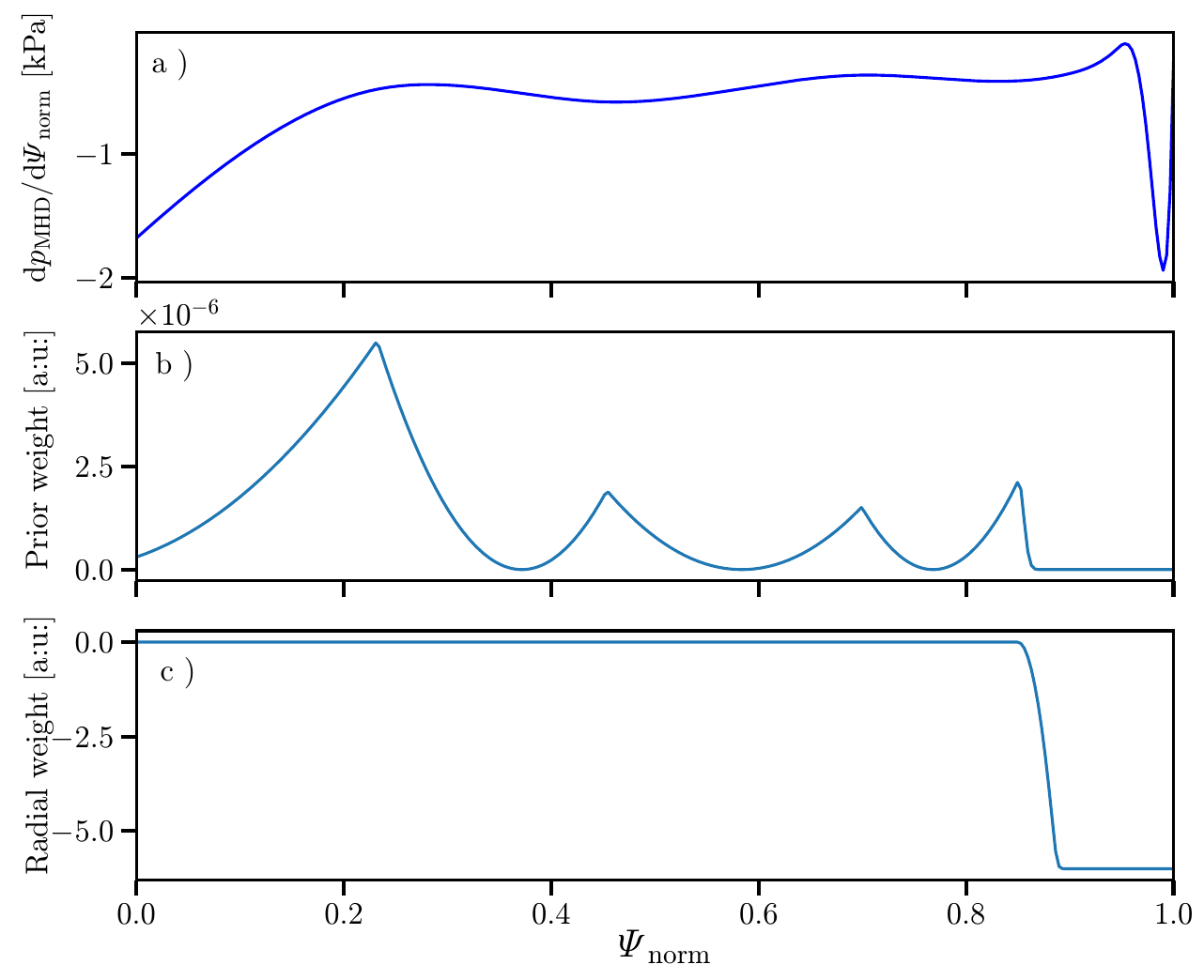}
    \caption{\label{fig:example_curvature_constraint} Illustration of the radially weighted curvature constraint for the $\mathrm{d} p/\mathrm{d}\Psi_\mathrm{norm}$ profile. a) shows the profile, which is the same as in \cref{fig:kinetic_reconstruction}. b) Shows the prior before it is integrated radially and c) shows the radial weight function included in b).}
\end{figure}

A log-normal constraint centered around a zero second derivative is assumed for all profiles except those that are represented as $\exp$(B-spline).
For those, the second derivative of the B-spline is considered directly before applying the exponential function.
A radially resolved weight function allows higher curvature near the edge to accommodate the pedestal.
As the weight function, this prior uses the sum-of-squares of the log-normal of the curvature value on the profile grid divided by the number of radial grid points.
\Cref{fig:example_curvature_constraint} shows an example for the curvature constraint for the $\mathrm{d} p/\mathrm{d}\Psi_\mathrm{norm}$ profile that is the solution of the kinetic reconstruction discussed in section \ref{sect:kinetic_reconstruction}. 
%Figure 4. a) shows the profile itself, b) shows the radially resolved curvature constraint and c) shows the radial weight function included in b). %TBA: duplicating the caption doesn't seem justified without providing more context or referring to the figures to make a point

\subsubsection{Positivity and monotonicity constraints}
Similar to the curvature constraint, the positivity and monotonicity constraints also use a radial grid of weights to allow for stronger influence, for example, near the plasma edge where the curvature constraint is relaxed.
They also allow for a sign to be supplied to enable switching between positivity and negativity constraints. 
The challenge with these two constraints is to facilitate a smooth transition between passive mode, e.g., the profile has the expected sign, and penalizing mode, e.g., the profile does not have the expected sign.
If this transition is not numerically differentiable, it will cause problems for the \gls{MAP} optimization.
To allow a smooth transition, we implemented the following with $y$ either the profile itself or its gradient times the expected sign: %TBA: not clear how length scale transitions are involved here, but if that's important it should be clarified in relation to the passive vs penalizing modes described above

\begin{align}
    p_\mathrm{prior} = 
    \begin{cases}
        \log(1 + e^{-\alpha y}) \text{ when y is positive}
        \\
        -\alpha \frac{y}{2} + \log(2.0) \text{ when y is negative}.
    \end{cases}
\end{align}
with $\alpha$ the typical, inverse scale length of $y$ defined manually.
Like the curvature constraint, this prior is also normalized by the number of radial points and goes into the posterior as a sum-of-squares.

\subsubsection{Parameter boundary constraint}
Since the inferred parameters use a transformation (see section \ref{sect:transformations}) that maps infinity in optimizer space to a finite number on the boundary of the physics space, it is necessary to enforce a prior on the optimization parameters $x$ directly.
While a simple log-normal distribution, e.g. \mbox{$\propto -x^n$} with $n=2$, might be the most obvious choice since it matches the transformation based on the cumulative normal distribution, we found in practice that using even order $n>2$ yields better results.
For the analysis shown in section \ref{sect:results} $n=6$ is used, which reduces the impact of the prior when at the center of the database, e.g., for $\vert x \vert < 1$, while the log-prior increases rapidly when approaching the edges. Of course, for the results shown in section \ref{sect:results}, we have the luxury of knowing that $\vert x \vert < 1$ is the ground truth because we are working with an artificially created scenario. However, since E-Forward-NN relies only on generated data, the training database for E-Forward-NN can be, in principle, extended indefinitely such that $\vert x \vert < 1$ is valid even for the diverse conditions of real experiments.

\subsection{The IDSServer}\label{sect:IDSServer}
The individual objects exchange information via the IDSServer database component that stores data internally according to the \gls{IMAS} schema. It also handles data storage and retrieval via the \gls{IMAS} library and transporting data across processes in case of parallel evaluations of the posterior.
Internally, the IDSServer does not use the \gls{IMAS} library directly because the nested nature of the data object is not well-suited for \gls{IDA}. For example we need \verb|thomson_scattering.channel.t_e.data| to be one memory contiguous array but in reality each \verb|channel.t_e.data| is an independent object that cannot be used in numpy operations. %TBA: can we provide a short explanation of why?
Instead, each quantity (e.g., \verb|thomson_scattering.channel.t_e_reconstructed|) is stored as a memory-contiguous tensor, which is essential for utilizing vectorization across many parameter samples.

\section{Simultaneous kinetic profile and magnetic equilibrium reconstruction for an \Gls{ITER} plasma}
\label{sect:results}

In this section, we will assemble an entire workflow for the equilibrium and profile reconstruction in several steps, starting with the simplest scenario: \gls{Te} and \gls{ne} reconstruction using \gls{TS} local measurements and the vibration-compensated phase from the \gls{TIP} and \gls{DIP} diagnostics \cite{van2017tests}.
A comprehensive list of the machine description files used for this case and all the other cases in this section is provided in \cref{tab:diag_MD}. As for the plasma scenario, a full-field D-T \Gls{ITER}-like discharge predicted by the FUSE\cite{meneghini2024fuse} code is used as the ground truth for artificial data generation.

\subsection{Artificial data generation}\label{sect:art_data}
Artificial data for these tests are generated by the synthetic diagnostics that make up the forward models in the IDA framework.
Hence, the ground truth is known for the cases discussed in this section, which allows a straightforward comparison of the results with the underlying truth.
The only exception is the magnetic measurements, and the flux matrix used for mapping the profile diagnostics, which are generated with predictive EFIT directly and not the \gls{NN}.
Since none of the synthetic diagnostics include models for systematic errors that can arise from calibration or any other issues expected in the real world, we can only add uncorrelated stochastic noise and not systematic uncertainties in the artificial data.
To effectively assess the accuracy of the quantities inferred by \gls{IDA}, more sophisticated synthetic diagnostics are needed that are capable of modeling realistic systematic uncertainties.

For the stochastic noise, we consider an average, relative noise level derived from the median signal of all diagnostic channels, a relative percentage noise for each individual channel of a diagnostic, and a flat constant noise.
The magnitude of the uncertainties is chosen arbitrarily except for the vibration-compensated phase and the polarimetry phase shift of \gls{TIP} and \gls{DIP}, for which we use the design specifications at the time of writing.
All noise sources are Gaussian.
\Cref{tab:diag_unc} shows the breakdown of the noise levels for all diagnostics considered in the \Gls{ITER} workflow.
At present, we only draw a single sample for the time slice considered and do not consider multiple data points for each measurement in the reconstruction of the profile+equilibrium.

\begin{table}[]
    \centering
    \begin{tabular}{|c|c|c|c|}
        \hline
        System Described & Root IDS & Shotnumber & Run ID\\
        \hline\hline
        wall shape & wall & 116000 & 4 \\
        \hline
        coils & pf\_active & 111001 & 203 \\
        \hline
        probes and flux loops & magnetics & 150100 & 4 \\
        \hline
        TS & thomson\_scattering & 150301 & 1 \\
        \hline
        TIP & interferometer & 150305 & 2 \\
        \hline
        DIP & interferometer & 150610 & 2 \\
        \hline
    \end{tabular}
    \caption{\label{tab:diag_MD} This table lists the \gls{IMAS} shot numbers and run IDs for all \Gls{ITER} machine descriptions used in this work. These were the latest available as of 02/2024.}
\end{table}

\begin{table}[]
    \centering
    \begin{tabular}{|c|c|c|c|c|}
        \hline
        Measurement & \# Signals & Average relative uncertainty & Relative uncertainty & Flat uncertainty\\
        \hline\hline
        coils & 11 & \SI{1}{\percent} & 0 & 0  \\
        \hline
        poloidal field probes& 795 & \SI{1}{\percent} & \SI{5}{\percent} & 0 \\
        \hline
        toroidal flux loops & 40 & \SI{1}{\percent} & \SI{5}{\percent} & 0 \\
        \hline
        plasma current & 1 & \SI{5}{\percent} & \SI{5}{\percent}  & 0 \\
        \hline
        Thomson scattering & 71 & \SI{5}{\percent} & 0  & 0 \\
        \hline
        interferometer & 7 & 0  & 0 & 10 degrees \\
        \hline
        polarimeter & 7 & 0  & 0 & 0.1 degree \\
        \hline
        ion pressure & 8 & \SI{5}{\percent} & 0  & 0 \\
        \hline
    \end{tabular}
    \caption{\label{tab:diag_unc} Artificial \Gls{ITER} data is generated by adding normally distributed noise with the uncertainties described in this table to a target case from our database.}
\end{table}
We use machine descriptions from the \verb|ITER_MD| machine description database for all diagnostics and other hardware components. The IDS were obtained in February 2024, so some might have been superseded by newer versions since then. \Cref{tab:diag_MD} summarizes all machine description files used to model \Gls{ITER} in this work.

\subsection{Inference of electron kinetic profiles}
\begin{figure}
    \includegraphics[width = \textwidth]{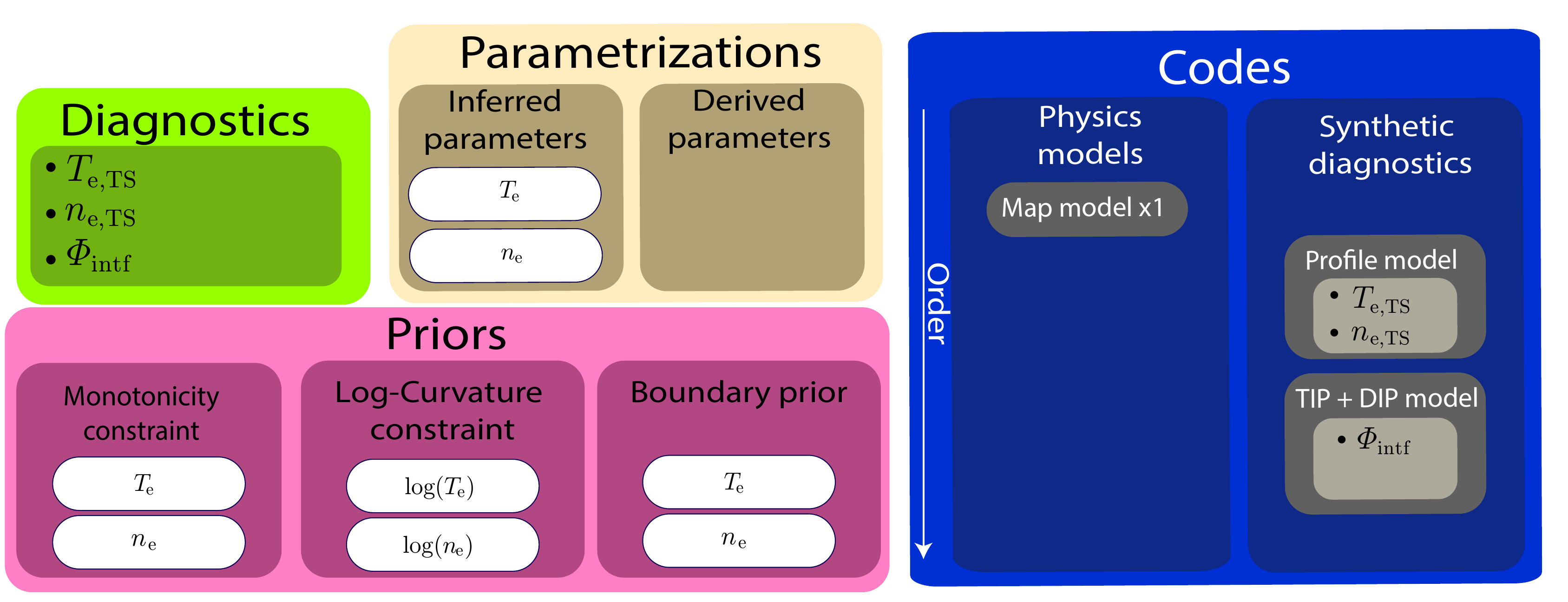}
    \caption{\label{fig:profile_reconstruction_illustration} Illustration of the various components used for the profile-inference. For detailed explanation of each component please refer to section \ref{sect:Components}.}
\end{figure}

We begin with the simplest scenario: inference of the \gls{Te} and \gls{ne} profiles constrained by \gls{TS} and the TIP and DIP interferometers.
Note that the geometry of edge \gls{TS} was not available in the \gls{IMAS} \Gls{ITER} machine description file and is, therefore, not included. 
\Cref{fig:profile_reconstruction_illustration} shows the specific components included in this scenario.
Only the temperature and density measurements at the \gls{TS} positions and the vibration-compensated phase of the TIP and DIP diagnostics are considered.
The \gls{Te} and \gls{ne} profiles are the only inferred parameters and the normalized flux matrix is loaded using the ground truth as the reference.
The code flow includes a single run of the mapping code that calculates $\Psi_\mathrm{norm}$ values for the spatial coordinates of the diagnostics.
This code is only run once before the optimization because the flux matrix is static in this scenario. During the inference, the profile code is run to update the \gls{Te} and \gls{ne} values for the diagnostics.
It is followed by the interferometer code that computes the vibration compensated phase for TIP and DIP $\phi_\mathrm{intf}$.
For priors, curvature constraints and monotonicity constraints are applied to the log profiles.
Even though this case does not include any \gls{NN}, we also include the boundary prior for \gls{Te} and \gls{ne} here for consistency with the other scenarios discussed in this section.

\begin{figure}
    \includegraphics[width = \textwidth]{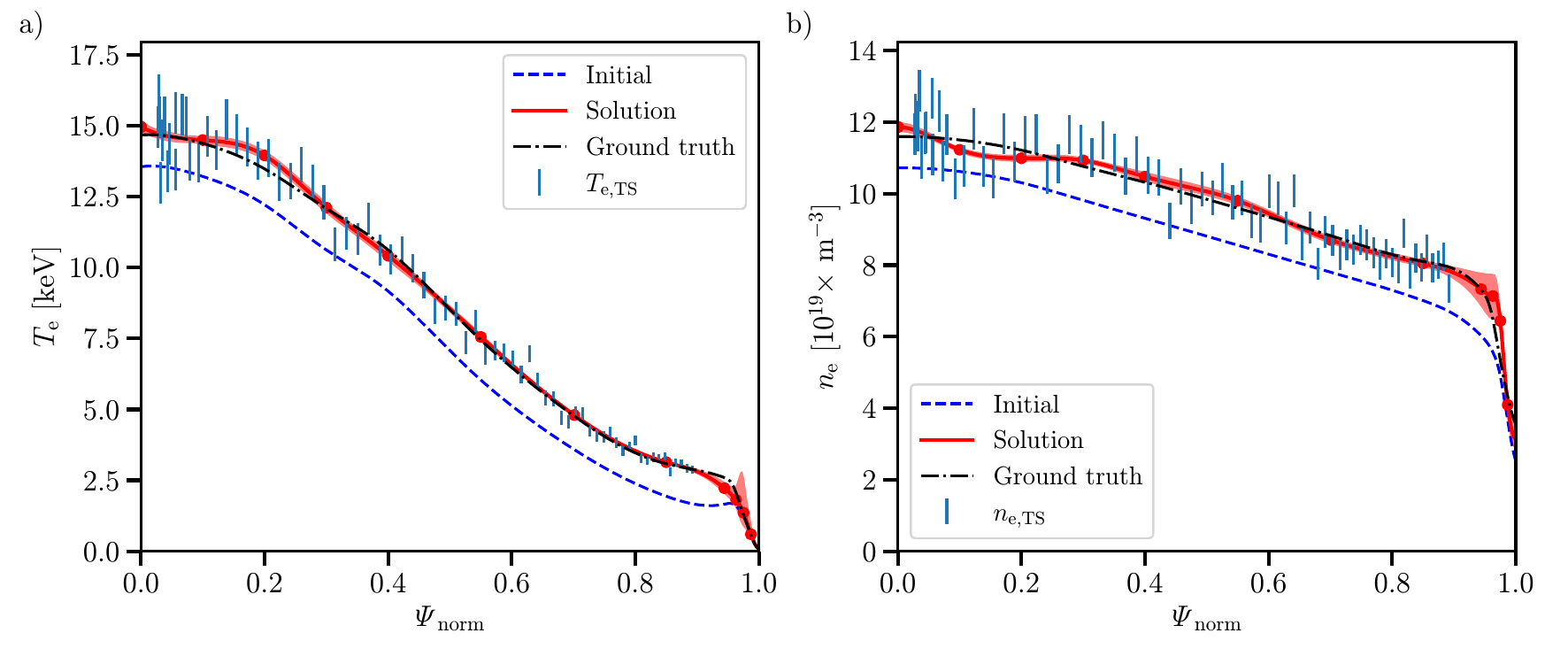}
    \caption{\label{fig:profile_reconstruction} Inferred posterior of  a) \gls{Te} and b) \gls{ne} are illustrated by red-shaded error-bands representing the 15-85 percentile of all samples. The red dots indicate the B-spline knots. The \gls{TS} measurements and their associated uncertainties are shown by vertical bars. The initial guess for the parameters where the \gls{MAP} starts is indicated by blue dashed lines and the ground truth is represented by black dot-dashed lines.}
\end{figure}

\begin{figure}
    \includegraphics[width = \textwidth]{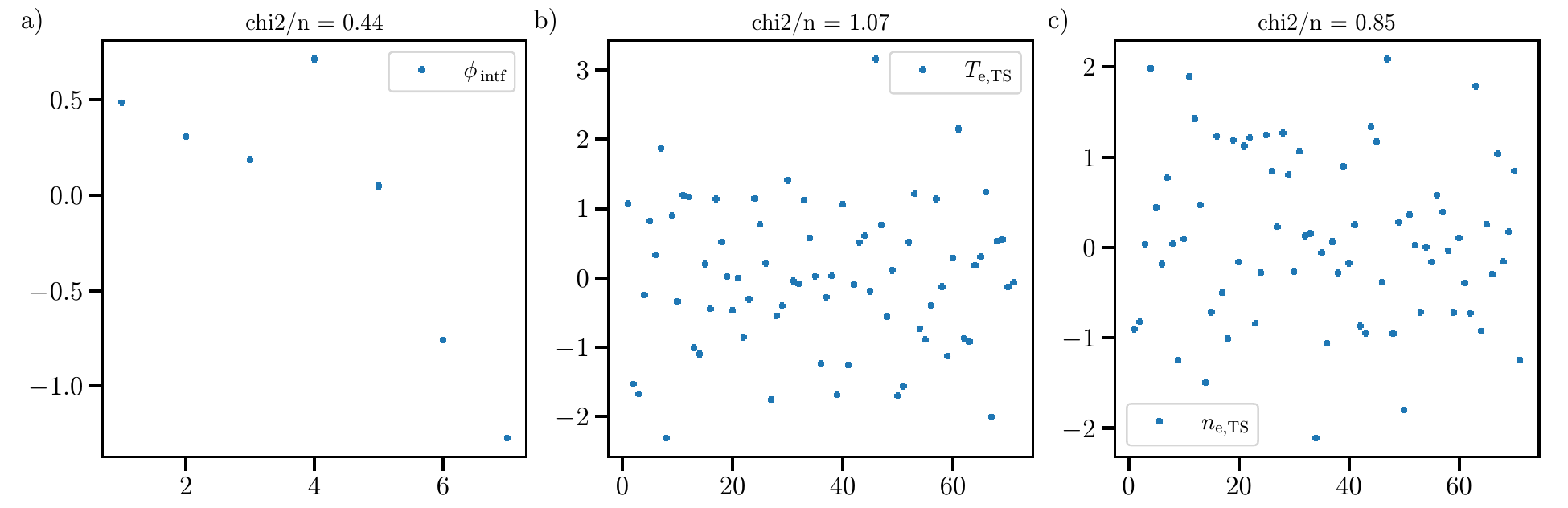}
    \caption{\label{fig:profile_reconstruction_residuals} Residuals normalized by uncertainty are plotted as a function of the diagnostic channel number (x-axis) for a) the vibration compensated phase of \gls{TIP} and \gls{DIP}, b) the \gls{Te} measurements and c) the \gls{ne} measurements of TS. The number above the plots indicates the $\chi^2$ divided by the number of channels, $n$. The residuals are close to one, indicating a good fit for \gls{Te}.}
\end{figure}

\Cref{fig:profile_reconstruction} shows the result of the profile inference with \gls{MAP} that took less than \SI{10}{\second} including uncertainty propagation with \num{10000} samples when parallelized on a 64-core machine.
The \gls{MAP} starts at parameters corresponding to the blue dashed profile.
The result (red solid line) and associated uncertainty (red shaded area enveloping \SI{70}{\percent} of all samples) is close to the ground truth (black dot-dashed line).
There are some discrepancies near the magnetics axis ($\Psi_\mathrm{N}=0.20$ for \gls{Te} and $\Psi_\mathrm{N}=0.18$ for \gls{ne}) where the stochastic error in the \gls{TS} data shows apparent systematic behavior.
It is also noticeable that a clear pedestal is visible even though the \gls{TS} data does not cover this region, particularly for \gls{Te} where the TIP and DIP diagnostics do not deliver any information.
The pedestal is an artifact introduced by the parameter transformation and the associated prior which only allows for little flexibility at the edge. % TBA: this sounds like it was accidental, which I think is underselling the priors
%Specifically, the outermost spline coefficient of the \gls{Te} profile has a minimal range in the training FUSE database, giving it little flexibility. % TBA: it includes physics!
Specifically, the outermost spline coefficient of the \gls{Te} profile has a minimal range in the FUSE database as a consequence of the stiff transport in H-mode pedestals. While the boundary prior is not strictly necessary since no neural network is used, it will be needed in the next two sections, so for consistency we also included it here.
\Cref{fig:profile_reconstruction_residuals} shows the residuals of the profile reconstruction and the $\chi^2$ normalized by the number of channels, $n$, for each diagnostic type.
Assuming that $n \gg N$ where N is a number of parameters, $\chi^2/n \approx 1$ for each diagnostic indicates a good fit. 
Since all $\chi^2/n$ are close to unity, the MAP solution is well-converged and fits the data within the uncertainty. 

\subsection{Bayesian equilibrium inference constrained by magnetic measurements only}\label{sect:mag_recon}

\begin{figure}[h!]
    \includegraphics[width = \textwidth]{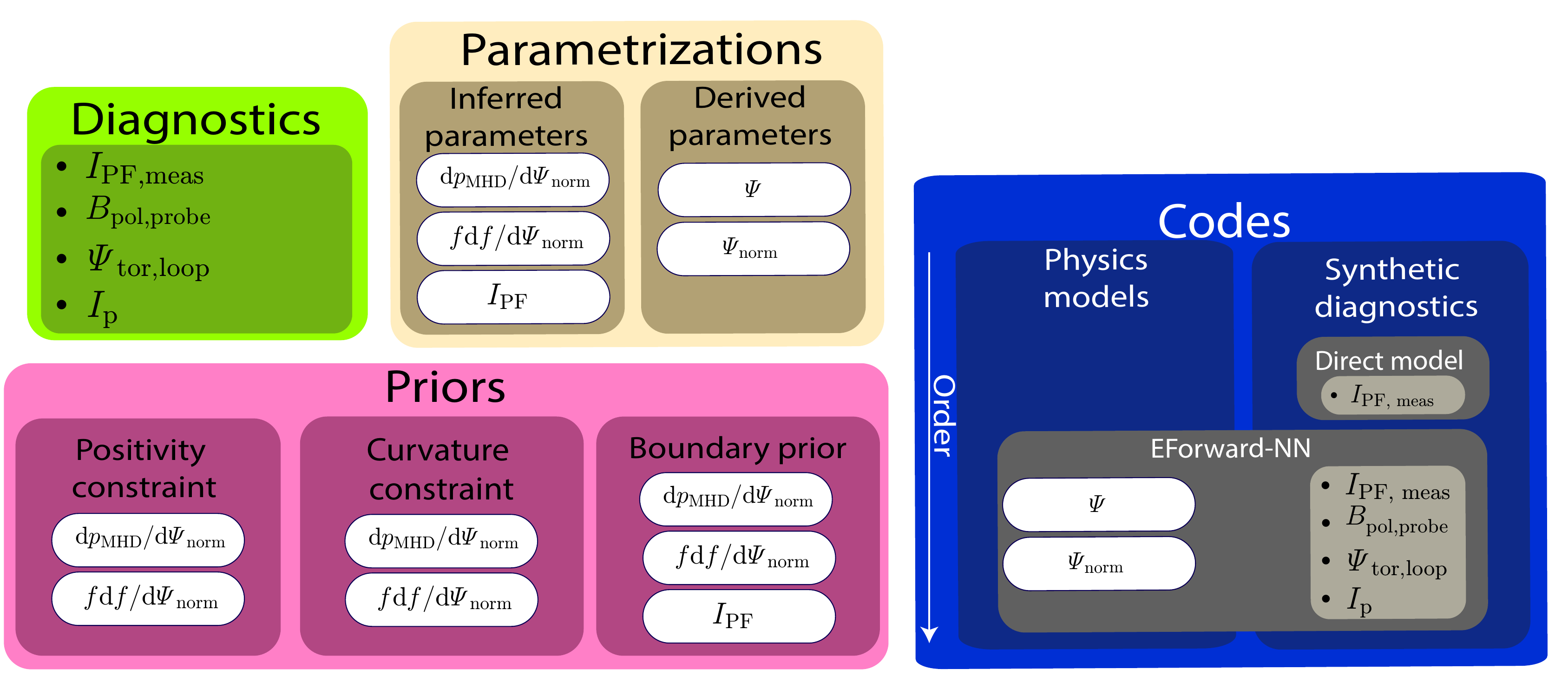}
    \caption{\label{fig:magnetic_reconstruction_illustration} Illustration of the various components considered in the inference of the equilibrium using external magnetic measurements only.}
\end{figure}

For the next step, we infer the equilibrium profiles using only external magnetic measurements. \Cref{fig:magnetic_reconstruction_illustration} shows the specific components included in this scenario. We infer the parametrized $p'$ and $ff'$ profiles together with the poloidal field currents $I_\mathrm{PF}$ using poloidal field probes, toroidal flux loops, the measured plasma current, and the measurement of the currents in the poloidal field coils.
For priors, we use curvature and negativity constraints for $p'$ and $ff'$ in conjunction with the boundary prior for all inferred parameters. Note that the negativity prior is not generally applicable for all scenarios, and it needs to be replaced with a physics informed prior in the future.
Before the optimization, the inferred $I_\mathrm{PF}$ currents are set to be equal to the mean of the measured current for each coil, which speeds up the optimization and makes it more robust. % TBA: why set these equal to the mean and not the measured currents in each coil? It seems like that would be closer to the final solution. Is it too biased??
The only code block used in this inference is E-Forward-NN, and it only evaluates its synthetic diagnostics. The flux matrices are only evaluated during the uncertainty propagation as a result, since they are not needed for the \gls{MAP} inference based on magnetic measurements alone.

\begin{figure}[h!]
    \includegraphics[width = \textwidth]{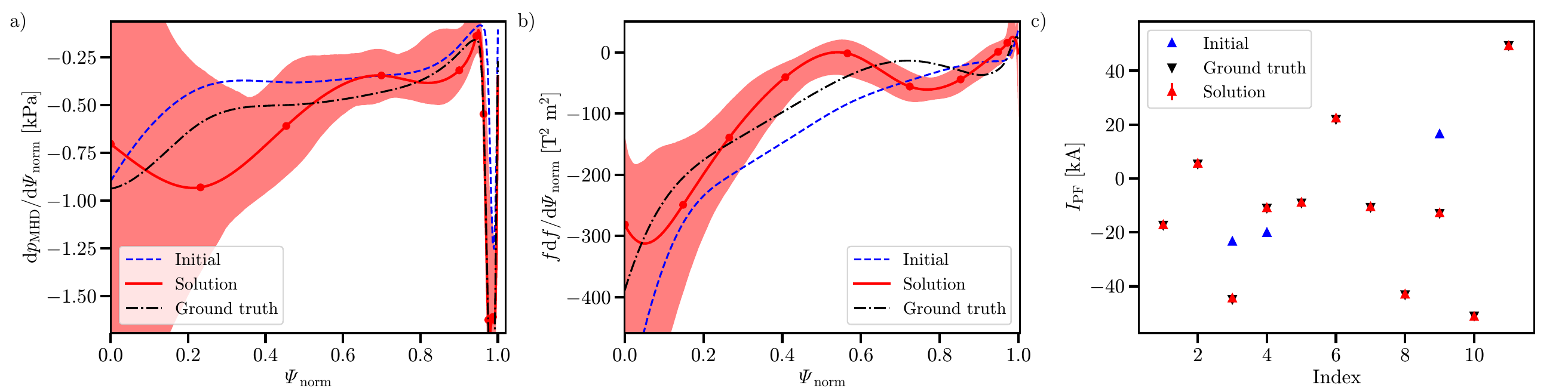}
    \caption{\label{fig:magnetic_reconstruction} Reconstruction of a) $p'$, b) $ff'$ and c) $I_\mathrm{PF}$. The red-shaded region indicates upper and lower error bars representing the 15-85 percentile of all samples. The red dots indicate the spline knots.}
\end{figure}

\begin{figure}[h!]
    \includegraphics[width = 0.8\textwidth]{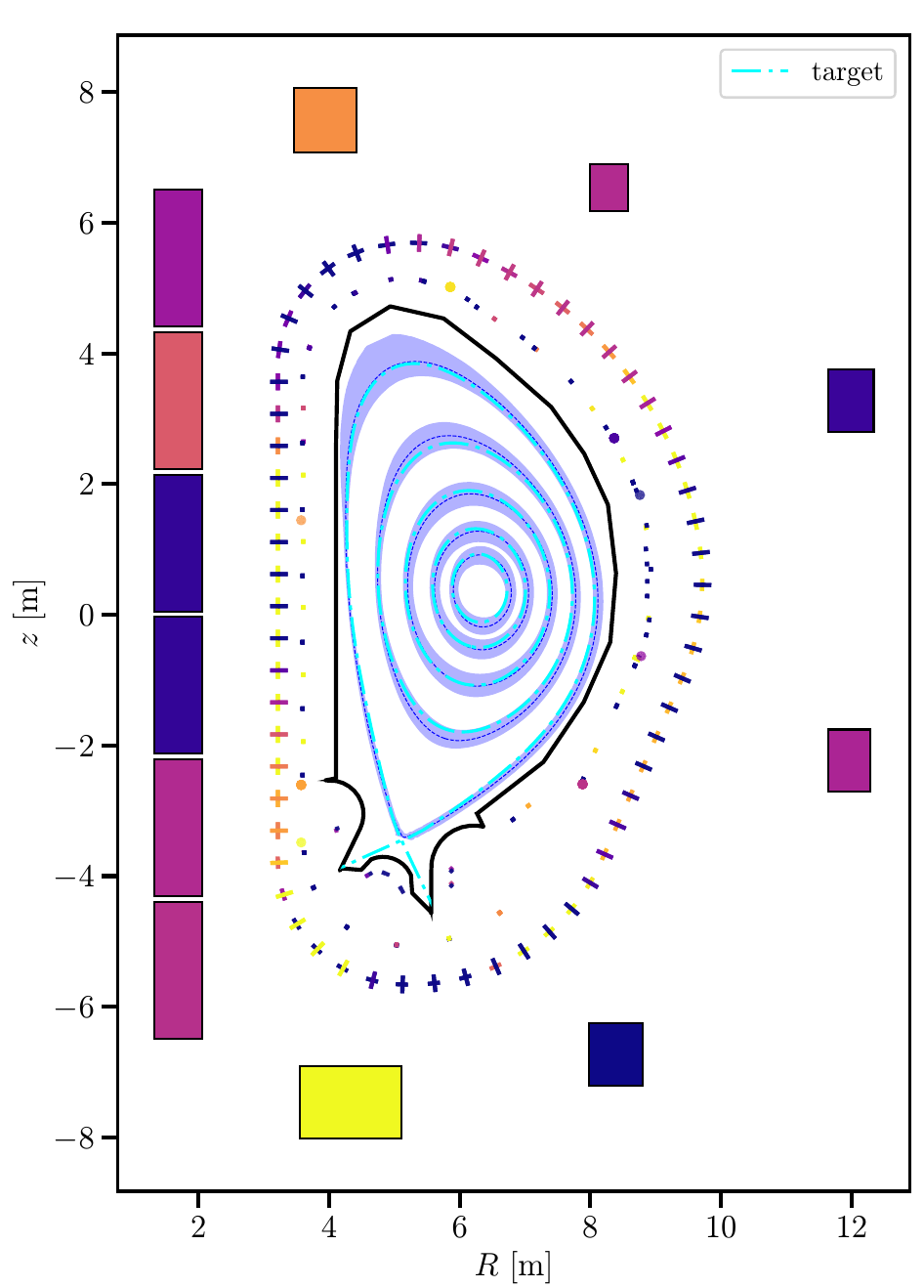}
    \caption{\label{fig:magnetic_reconstruction_flux} Flux contours of the magnetic equilibrium are indicated by blue dashed lines. The cyan dot-dashed lines indicate the contour lines of the ground truth. The blue shaded area indicates the uncertainty of the flux matrix representing the 15-85 percentile of all samples.}
\end{figure}

\Cref{fig:magnetic_reconstruction} shows the parameters inferred with \gls{MAP}.
This took about one \SI{1}{\minute} to complete, most of which (\SI{50}{\second}) is spent calculating the error bars for the flux surface contours.
Compared to the \gls{Te}/\gls{ne} reconstruction (see \cref{fig:profile_reconstruction}), the uncertainties for $p'$ and $ff'$ are rather large. This is not surprising since the external magnetic measurements imply little information about the plasma core. \Cref{fig:magnetic_reconstruction_flux} shows the contour lines of the flux matrix and their associated uncertainties propagated from the uncertainties of the spline coefficients and $I_\mathrm{PF}$ currents. As expected from the large uncertainties in the profiles, the uncertainty of the flux surface contours is also quite large.

\begin{figure}[h!]
    \includegraphics[width = \textwidth]{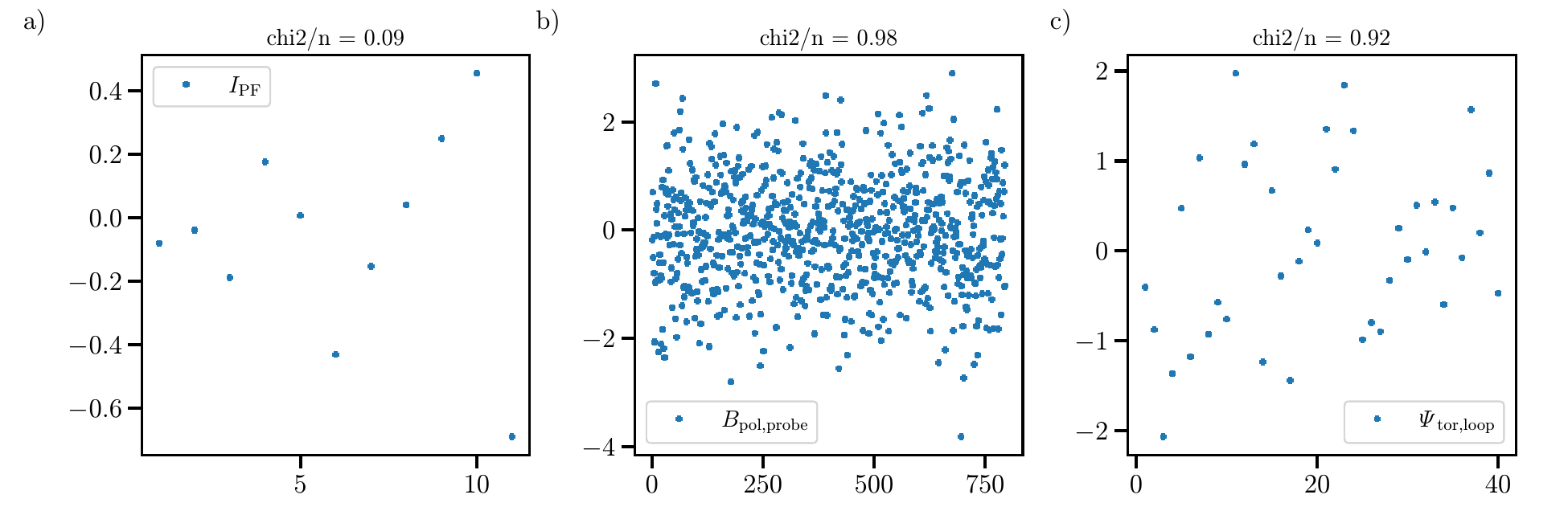}
    \caption{\label{fig:magnetic_reconstruction_residuals} Residuals for a) the measured poloidal field current, b) the poloidal field probes and c) the toroidal flux loops. The number above each indicates the $\chi^2$ divided by the number of channels, $n$. % The poloidal field currents seem to be quite strongly overfit, while the $\chi^2/n$ for the probes and flux loops is close to one.
    }
\end{figure}

The residuals of the \gls{MAP} estimate are shown in \cref{fig:magnetic_reconstruction_residuals}. Not depicted is the fit of the plasma current, which has a $\chi^2=0.0$ indicating that $I_p$ is overfitted and other diagnostics do not provide contradictory information.  %this also means that there is no redundant measurement for Ip, which is weird, since poloidal B probes depend on Ip. Once we will include a partial rogowsky coils, it will make more sense. 
The other quantities seem to be fit well with $\chi^2/n \approx 1$.

While the comparisons described in this section have some similarity to that described in ref. \citenum{lao1985efit}, they are not equivalent. 
The three primary differences are 1) we add noise to the artificial data being fit, 2) the reference case is a equilibrium that includes more information than can be extracted from external measurements alone (such as the pedestal and boostrap currents), and 3) $p'$ and $ff'$ profiles are represented by B-splines with 12 knots. 
As a result, this comparison is more representative of an experimental reconstruction and demonstrates that Bayesian \gls{IDA} can provide useful solutions for data-poor inverse problems.

\subsection{Kinetic equilibrium reconstruction}\label{sect:kinetic_reconstruction}
\begin{figure}[h!]
    \includegraphics[width = \textwidth]{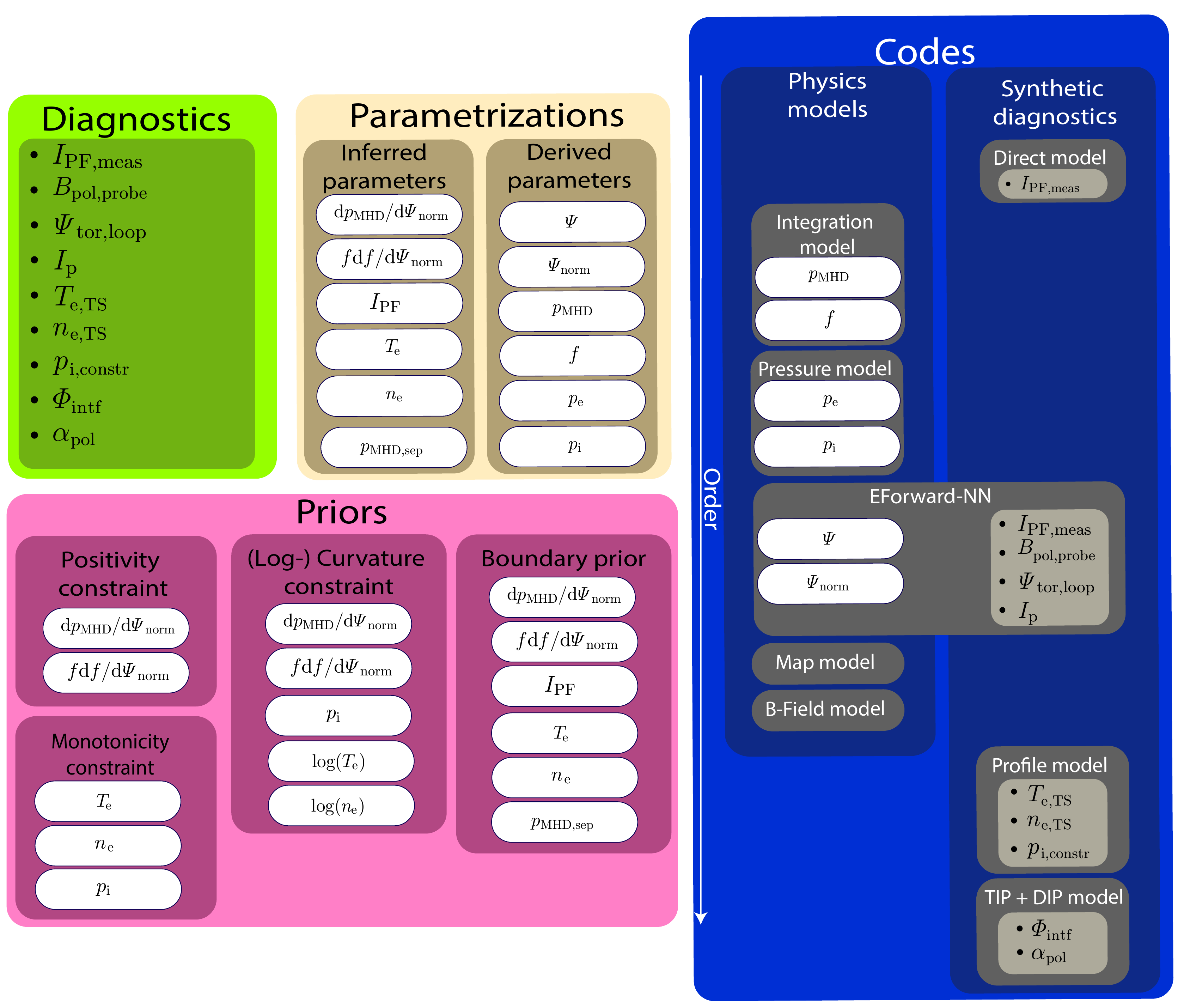}
    \caption{\label{fig:full_reconstruction_illustration} Illustration of the various components considered in the combined inference of kinetic profiles and the magnetic equilibrium.}
\end{figure}

Finally, the kinetic equilibrium reconstruction combines the two previous discussed cases. It also includes an additional diagnostic, the polarimeter phase $\phi_\mathrm{pol}$ measured by TIP and DIP, and an additional inferred parameter, the magnetic pressure at the separatrix $p_\mathrm{MHD,sep}$. \Cref{fig:full_reconstruction_illustration} shows the overview of all the components considered. This workflow combines all components of \cref{fig:profile_reconstruction_illustration} and \cref{fig:magnetic_reconstruction_illustration}, and adds the following derived parametrizations: 

\begin{enumerate}
    \item $p_\mathrm{MHD}$ represented by a B-spline as $\log(p_\mathrm{MHD})$
    \item $f$ represented by a B-spline directly
    \item $p_\mathrm{ion}$ the total ion pressure represented as a B-spline directly %TBA: why not use a log for this? It isn't intuitive so it might be worth explaining. It is kinda complicated
    \item $p_\mathrm{e}$ represented by a B-spline as $\log(p_\mathrm{e})$
\end{enumerate}

It also introduces three new codes:
\begin{enumerate}
    \item The profile integration model to calculate $p_\mathrm{MHD}$ and the diamagnetic function $f$
    \item The pressure model that calculates $p_\mathrm{e}$ from \gls{Te} and \gls{ne} to derive the ion-pressure with $p_\mathrm{i} = p_\mathrm{MHD} - p_\mathrm{e}$
    \item The B-field-model that derives the magnetic field vector from $\Psi$ and $f$ on the \gls{LOS} of the \gls{TIP} and \gls{DIP} diagnostics that are needed to calculate $\phi_\mathrm{pol}$.
\end{enumerate}

and two new measurements: 
\begin{enumerate}
    \item The polarimeter phase shift $\phi_\mathrm{pol}$
    \item The ion-pressure measurements $p_\mathrm{i,meas}$
\end{enumerate}
Since synthetic diagnostics for ions were not implemented yet, we added eight artificial measurements of the ion pressure profile. In future efforts, it will need to be replaced by actual ion diagnostics and predictive modeling for the fast ion profiles. Without this constraint or additional prior information about the ions the ion pressure would be completely free, and the electron pressure information could not be used effectively.
The only new priors compared to \cref{fig:profile_reconstruction_illustration} and \cref{fig:magnetic_reconstruction_illustration} are curvature and monotonicity constraints for $p_\mathrm{i}$ and a boundary prior for $p_\mathrm{MHD,sep}$.

\begin{figure}[h!]
    \includegraphics[width = \textwidth]{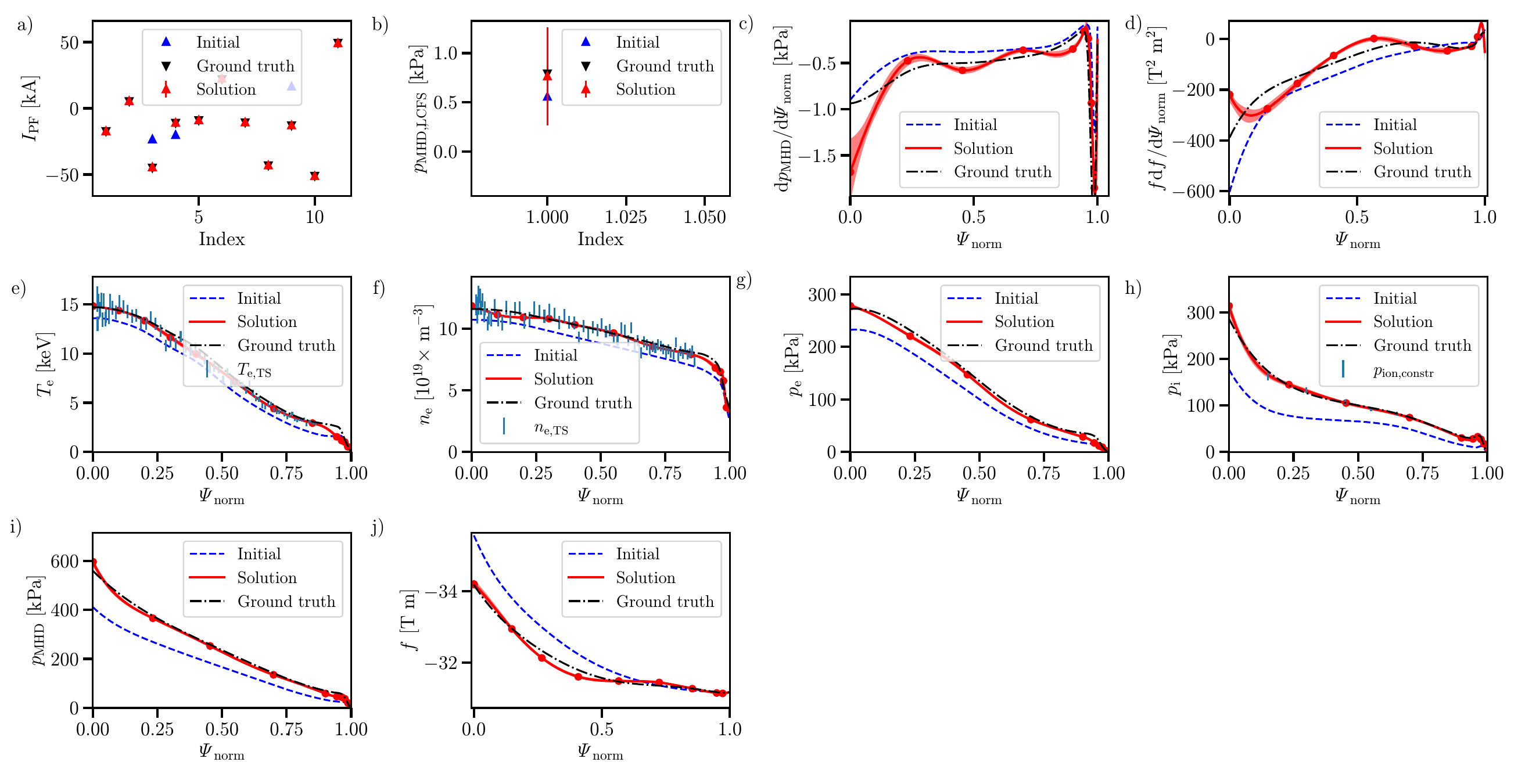}
    \caption{\label{fig:kinetic_reconstruction} Inferred a) $I_\mathrm{PF}$, b) separatrix pressure, c) $p'$, d) $ff'$, e) \gls{Te}, f) \gls{ne}, g) electron pressure $p_\mathrm{e}$, h) ion pressure $p_\mathrm{i}$ and its artificial constraints, i) the total MHD pressure $p_\mathrm{MHD}$,  and j) diamagnetic function $f$. The red-shaded region indicates upper and lower error bars representing 15-85 percentile of all samples. The red dots indicate the spline knots.}
\end{figure}

\begin{figure}[h!]
    \includegraphics[width = 0.8\textwidth]{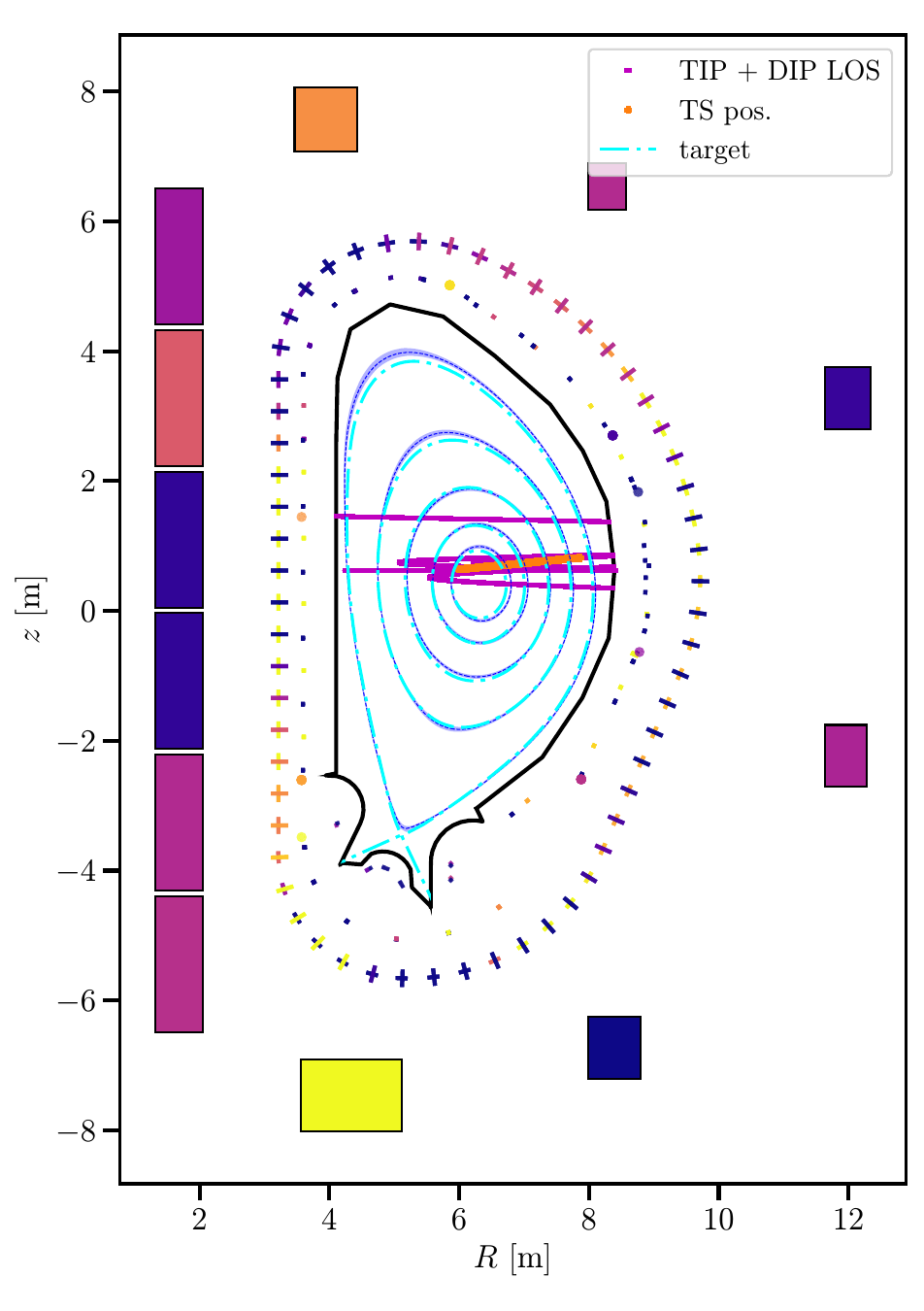}
    \caption{\label{fig:kinetic_reconstruction_flux} Flux contours of the magnetic equilibrium are indicated by dashed blue lines. The cyan dot-dashed lines indicate the contour lines of the ground truth. The blue shaded area indicates the uncertainty of the flux matrix representing the 15-85 percentile of all samples.}
\end{figure}

\Cref{fig:kinetic_reconstruction} shows the directly inferred and derived profiles and their uncertainties. This \gls{MAP} analysis took about \SI{3}{\minute} with uncertainty propagation.
Notably, the uncertainties for $p'$ and $ff'$ have been significantly reduced when compared to \cref{fig:magnetic_reconstruction}. This has a direct effect on the uncertainties of the flux matrix as shown by \cref{fig:kinetic_reconstruction_flux}, which also depicts the measurement positions of the \gls{TS} system and the lines of sight of the \gls{TIP} and \gls{DIP} diagnostics.
Compared to the case with magnetic measurements only (c.f. \cref{fig:magnetic_reconstruction_flux}) the uncertainty bands of the flux surfaces are much smaller.

\begin{figure}[h!]
    \includegraphics[width = \textwidth]{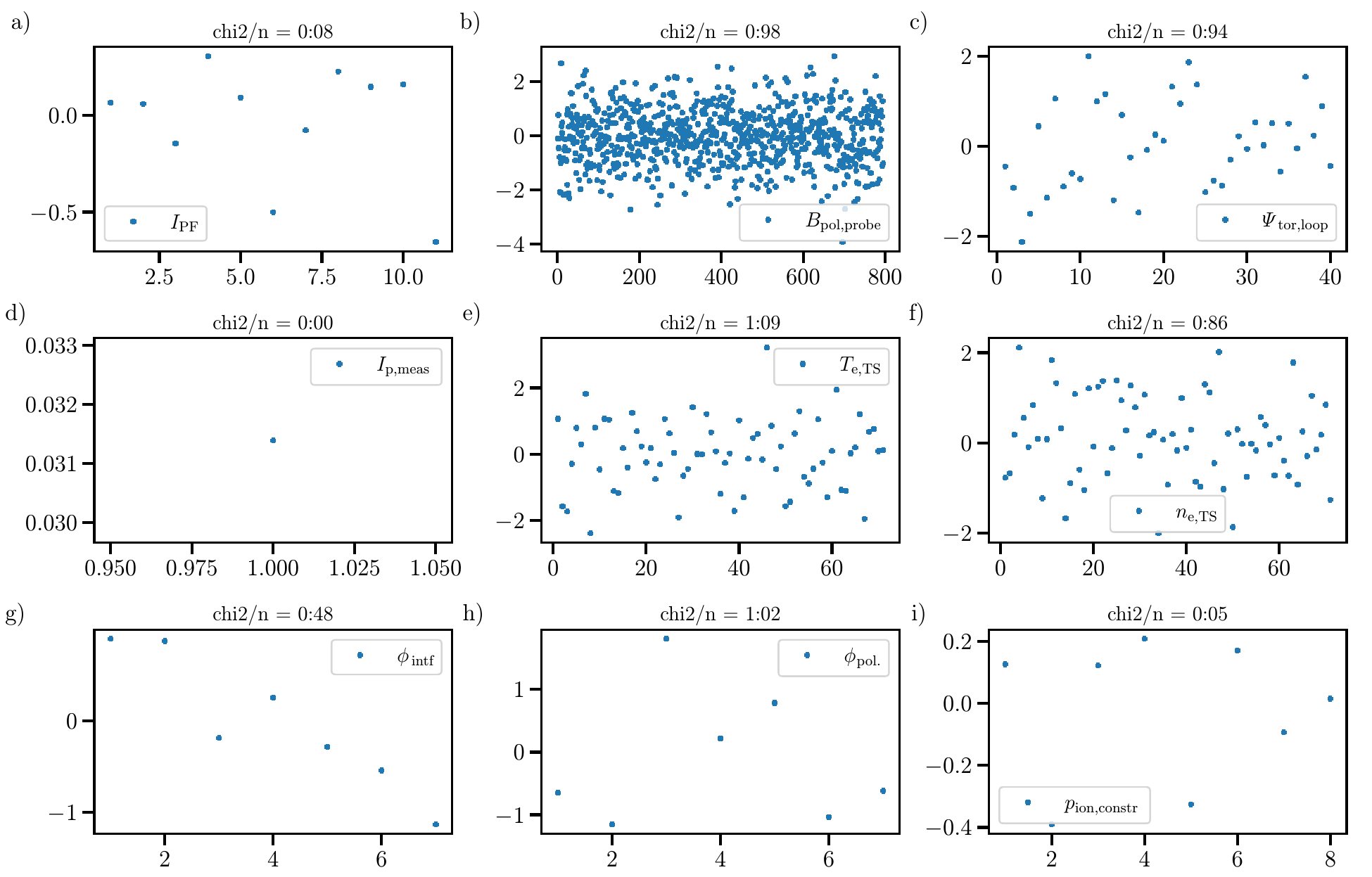}
    \caption{\label{fig:kinetic_reconstruction_residuals} Residuals for the a) measured poloidal field coil currents, b) poloidal field probes, c) toroidal flux loops, d) total plasma current e)\gls{Te} measured by \gls{TS}, f) \gls{ne} measured by \gls{TS}, g) vibration-compensated phase measured by the \gls{TIP} and \gls{DIP}, h) polarimeter phase shift rotation angle $\phi_\mathrm{pol}$, and finally i) ion pressure constraints. The number above indicates the $\chi^2$ divided by the number of channels, $n$.}
\end{figure}

\Cref{fig:kinetic_reconstruction_residuals} summarizes the residuals of the various diagnostics considered. Again, the total plasma current, $I_p$, (not depicted) is overfit with a $\chi^2=0.0$. All other residuals are close to one, indicating a good fit.

\section{Verification of MAP results\label{sect:Verification}}
While \gls{MAP} is computationally efficient it requires the posterior to be approximately a multivariate normal distribution near its maximum for the uncertainty propagation to be valid.
This is not guaranteed for a non-linear problem like equilibrium reconstruction. Unfortunately, more accurate alternatives to \gls{MAP}, such as \gls{MCMC} or nested sampling, are too computationally expensive for routine analysis. Nevertheless, we can use \gls{MCMC} to verify the accuracy of our \gls{MAP} results.

\Cref{tab:runtime} summarizes the runtime for the three workflows discussed in the previous section, broken down into individual steps, the MAP inference itself, the propagation of uncertainties using 10 kilosamples (kS) and the calculation of percentiles from the 10 kS. It also includes statistics of the MCMC analysis broken down into the sampling rate in kilosamples per second (kS\si{\per\second}), the effective sampling rate, i.e., sampling rate divided by correlation time (S\si{\per\second}) and total number of samples needed for convergence in megasamples (MS). The timings are based on a RHEL8 Linux machine with 2 x AMD EPYC 7513 32-Core processors and \SI{512}{\giga\byte} of RAM. Case A refers to the profile inference, B to the magnetic equilibrium reconstruction, and C to the kinetic equilibrium reconstruction.
\begin{table}[]
    \centering
    \begin{tabular}{|c|c|c|c|c|c|c|}
        \hline
        Case & MAP& Calculate& Calculate & MCMC sampling & Effective MCMC& Total\\
         &  inference [s] &  10 kS [s] & percentiles [s]  &  rate [kS \si{\per\second}] & sampling rate [S \si{\per\second}] & samples [MS]\\ % TBA: could more of these be converted to the same units for easier comparison? e.g. rates for MAP or total times for MCMC. Currently this requires the reader to do some homework. 
        \hline
        A & 0.7 & 1.24 & 1.49 & 17.0 & 27 & 200\\
        \hline
        B &1.6 & 54 & 27 & 20.0 & 22& 600\\
        \hline
        C & 37.6 & 81.3 & 40.3 & 2.46 & 1.8& 550 \\
        \hline
    \end{tabular}
    \caption{\label{tab:runtime} This table lists the performance metrics for the various calculations in the \gls{IDA} framework.}
\end{table}

\subsection{Profiles-only case}
First, we check the profiles-only case for which the posterior should be closest to a multivariate normal because it is well-constrained by measurements with normally distributed uncertainties. The EMCEE ensemble sampler is used with a move mixture of \SI{80}{\percent} differential evolution moves \cite{nelson2013run} and \SI{20}{\percent} snooker proposals \cite{ter2008differential} to test the posterior for multiple modes. A total of \num{1000} walkers each taking \num{200000} MCMC steps, i.e. 200 MS total, are needed to reach relative changes of the autocorrelation time, $\tau$, smaller than \SI{1}{\percent}. The first quarter of all MCMC steps are discarded as burn-in and the chain is thinned by $4\times \tau \approx 656$. From the resulting \num{225000} total samples \num{10000} samples are selected to be processed in the same manner as the samples drawn from a multivariate normal in \gls{MAP} analysis (see section \ref{sect:uncertainty_propagation}).

\begin{figure}
    \includegraphics[width = 0.8\textwidth]{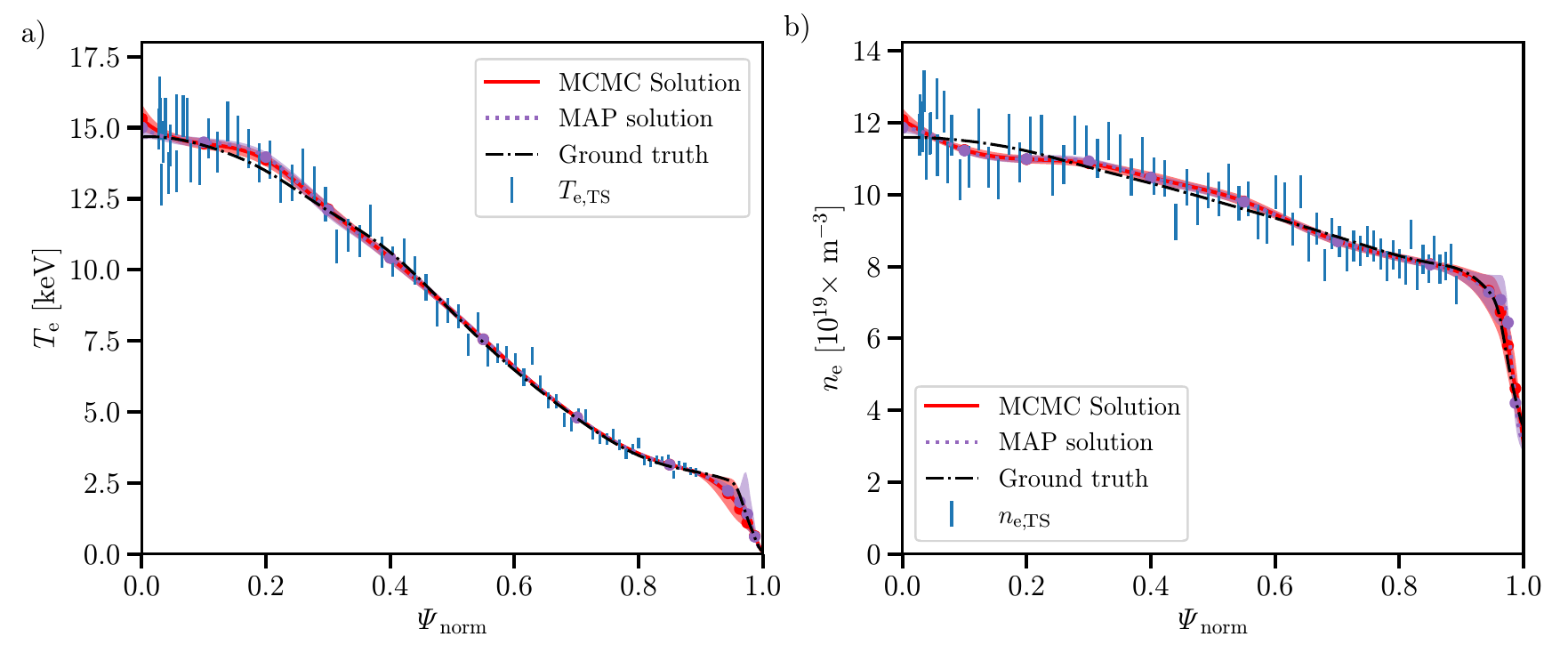}
    \caption{\label{fig:profile_reconstruction_MCMC} Reconstruction of a) \gls{Te} and b) \gls{ne}. The red-shaded region indicates upper and lower error bars representing the 15-85 percentile derived from \num{10000} MCMC samples. The purple line and shaded area indicate the results from the \gls{MAP} analysis.}
\end{figure}

\Cref{fig:profile_reconstruction_MCMC} shows the comparison of the \gls{MAP} results (purple) and the \gls{MCMC} results (red). The median of the samples is used to express the expected value for \gls{MCMC}. Notably, the log-posterior is larger for the \gls{MAP} (\num{-70.5}) than for the median parameters (\num{-71.1}), indicating that \gls{MAP} has likely been able to find the global maximum. The only visible distinctions occur near the edge where there is no \gls{TS} data, and the analysis is purely based on priors. The discrepancy there is not surprising as the priors cannot be described by a multivariate normal. For example, the boundary prior treats each parameter independently. Hence, the parameter space contours are rectangles while they would be ellipses for a multivariate normal. 
To summarize, the agreement is quite good, especially in data-rich regions.

\subsection{Magnetic measurements only case}
The \gls{MAP} analysis predicted significant uncertainties for $p'$ and $ff'$ because their determination is data-poor. We also expect the posterior of this problem to be less like a multivariate normal than the previous case because the synthetic magnetic measurements are non-linear with respect to the inferred parameters. Since \gls{MAP} extrapolates uncertainties of the posterior using the Hessian at the solution, we expect that the accuracy of the \gls{MAP} uncertainties will get worse because the non-linearity of the model becomes more apparent. To summarize, we expect MAP to show worse performance than MCMC, compared to the previous case. % TBA: This is the only case where you don't compare the log posterior magnitudes. How come?

\begin{figure}[!htbp]
    \includegraphics[width = \textwidth]{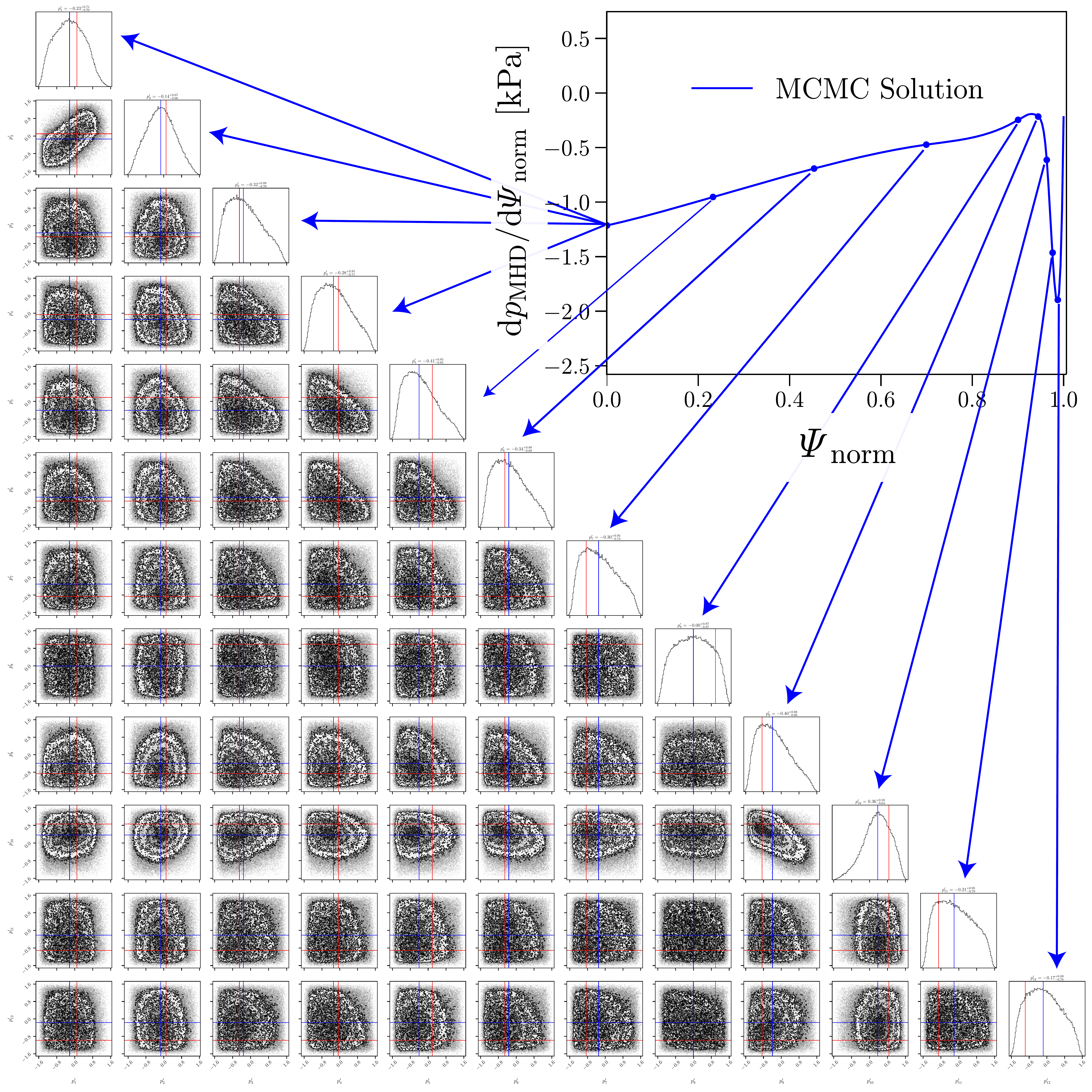}
    \caption{\label{fig:magnetic_reconstruction_corner} Corner plot for the spline coefficients of $p_{1}'$ to $p_{12}'$ in optimizer space. The blue vertical and horizontal lines indicate the median value derived from the \gls{MCMC} samples, and the red lines represent the mode provided by the \gls{MAP} analysis. The MCMC solution in the upper right provides context where each of the spline coefficients is located spatial. Note that the B-splines have four knots on the axis, hence there are also four spline coefficients.} %TBA: You should probably explain somewhere why there are 4 coefficients connected to the first knot...
\end{figure}

Similar to the profiles-only case, \num{1000} walkers are used, and \num{600000} \gls{MCMC} steps are needed for convergence yielding 600 MS total. Once the first quarter of samples is removed, and the chain is thinned, \num{152000} samples remain. We use these samples to create \cref{fig:magnetic_reconstruction_corner}, a corner plot \cite{corner} for the $p'$ profile.
It shows the 1-dimensional marginalized posterior as a function of each of the twelve spline coefficients (along the diagonal) and the marginalized posterior contours for pairs of coefficients (at the intersections within the table).
The red line indicates the \gls{MAP} solution, and the blue line indicates the median of the samples.
In both the 1-D and 2-D plots, there are clear features that are not described by a multivariate normal, such as the non-elliptic contours in 2D plots and the skewness in most of the 1D plots.

\begin{figure}
    \includegraphics[width = 0.8\textwidth]{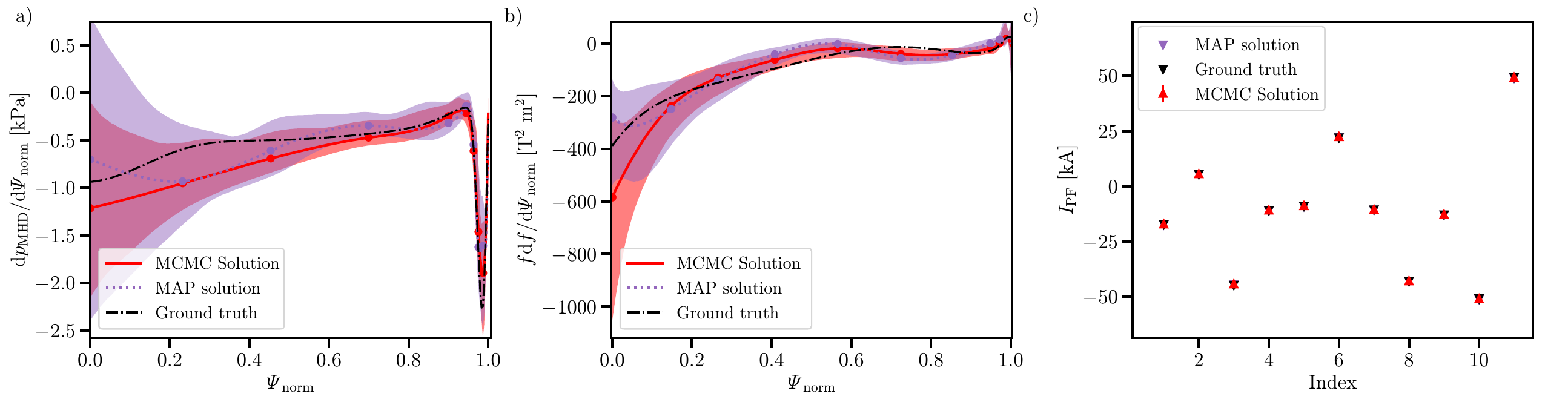}
    \caption{\label{fig:magnetic_reconstruction_MCMC}Reconstruction of the a) $p'$ profile, b) $ff'$ profile and c) $I_\mathrm{PF}$. The red-shaded region indicates upper and lower error bars representing 15-85 percentile derived from \num{10000} MCMC samples. The purple line and shaded area indicate the results from the \gls{MAP} analysis.}
\end{figure}

For the direct comparison of the \gls{MCMC} and \gls{MAP} uncertainty bands for parametrizations, the chain is thinned again to \num{10000} samples. \Cref{fig:magnetic_reconstruction_MCMC} shows the comparison between the \gls{MAP} and \gls{MCMC} results for the magnetic measurement only reconstruction. The median of the samples is used to express the expected value of the \gls{MCMC} solution. There is no discernible difference for $I_\mathrm{PF}$, which is unsurprising since the coil currents are strongly tied to the normal distributed measurements of the coil currents, resulting in the posterior being normally distributed with respect to the coil currents. However, as expected from the analysis of the corner plots, there are significant differences in the profiles. \gls{MAP} overestimates the uncertainties for both $p'$ and $ff'$ by more than \SI{50}{percent} in some regions, and we also see a difference in the expected values. While the \gls{MAP} uncertainties would certainly be better than nothing, here \gls{MCMC} is necessary to get meaningful uncertainties. 

\subsection{Kinetic equilibrium reconstruction}
\begin{figure}[!htbp]
    \includegraphics[width = \textwidth]{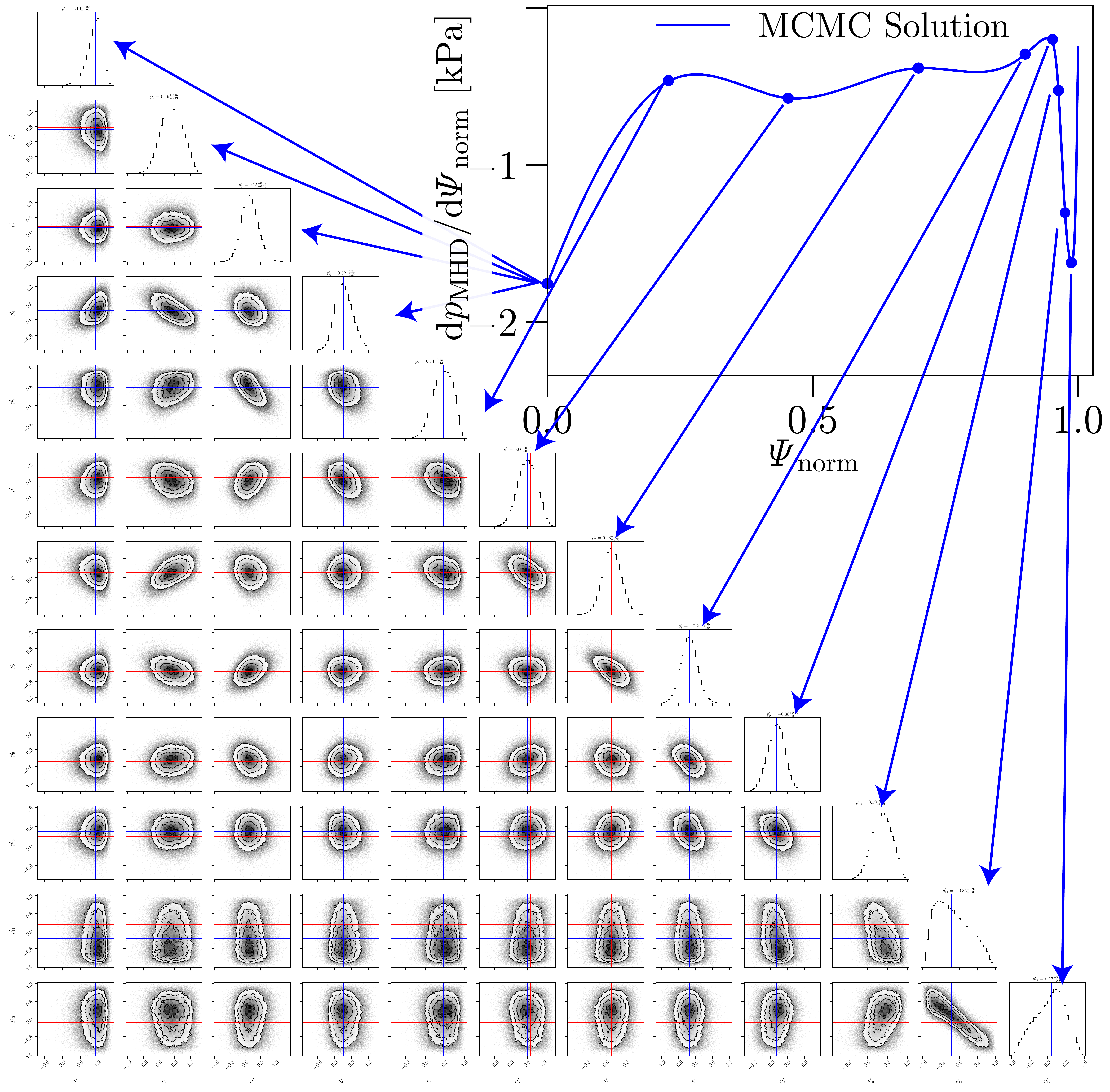}
    \caption{\label{fig:kinetic_reconstruction_corner} Corner plot for the kinetic equilibrium reconstruction. See \cref{fig:magnetic_reconstruction_corner} for a detailed description.}
\end{figure}

For the kinetic equilibrium reconstruction, we use the same recipe as before for the \gls{MCMC} analysis. This case is much more computationally expensive and full analysis takes several days to get a converged chain with \num{550} MS total, of which only \num{88000} samples remain after removing burn-in and thinning.
Again, the \gls{MAP} solution is likely the global maximum because the log-posterior (\num{-496}) is larger than the median of the \gls{MCMC} solution (\num{-508}).
Before we compare the error-bands, we revisit the corner plot for the $p'$ B-spline coefficients, that are shown in \cref{fig:kinetic_reconstruction_corner}. Unlike the previous case, the marginalized posterior is now quite close to a multivariate normal with only some residual skewness in some of the coefficients. Furthermore, the MAP solution (red lines) is close to the MCMC median (blue lines).

\begin{figure}
    \includegraphics[width = \textwidth]{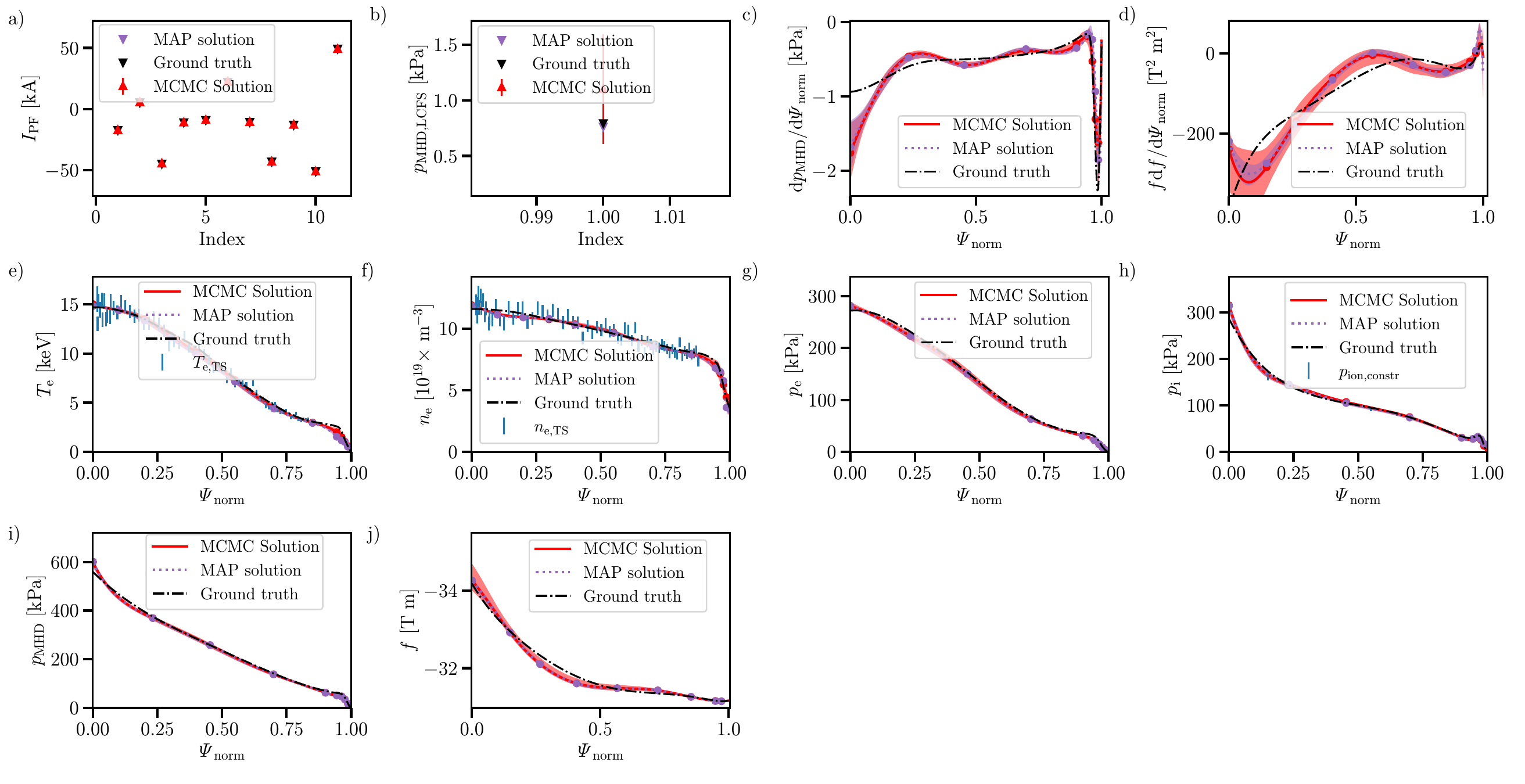}
    \caption{\label{fig:kinetic_reconstruction_MCMC} Inferred profiles similar to \cref{fig:kinetic_reconstruction} but the \gls{MCMC} median and propagated percentiles are shown in red. It is compared to the \gls{MAP} solution depicted in purple, and the initial state has been removed to reduce clutter.}
\end{figure}

\Cref{fig:kinetic_reconstruction_MCMC} shows the comparison of the parametrizations just as for the previous two cases. As expected from the corner plot, the uncertainty estimates for the $p'$ profile from \gls{MAP} are in good agreement with the \gls{MCMC} results. All other profiles show similarly good agreement, with the exception of $ff'$ for which \gls{MCMC} shows larger uncertainties.
This is not unexpected since there are no measurements that radially resolve $ff'$, and all the 
information on it comes from the flux surface shape information that external magnetic measurements, \gls{TS}, \gls{TIP}, and \gls{DIP} provide.
Since there is a non-linear relationship between these measurements and $ff'$, the posterior is not well approximated by a Gaussian, and the uncertainties estimated by \gls{MAP} are somewhat inaccurate compared to \gls{MCMC}.
The inclusion of additional diagnostics like the Motional Stark Effect diagnostic \cite{uzun2024designing} or the poloidal polarimetry system \cite{PoPola} are expected to improve the \gls{MAP} estimate and bring the uncertainties closer to the \gls{MCMC} results.

\section{Discussion and summary}\label{sect:Discussion}
We have developed a Bayesian inference framework that is capable of reconstructing self-consistent profiles and equilibria and their uncertainties in a single step in less than \SI{3}{\minute} on a single server node. To ensure that \gls{MAP} delivers reasonable uncertainties for this complex problem, we have compared its uncertainty estimates to that of \gls{MCMC} calculations and found good agreement for plasma parameters that are well-constrained by measurements.

The inference can be trivially parallelized across time points, and with sufficient resources, it can be faster than similar workflows like CAKE (see ref. \citenum{smith2024xloop} and references therein) that separate the profile inference from the equilibrium reconstruction and do not produce self-consistent uncertainties for all inferred quantities. With the computational capabilities provided by the \gls{IRI}, it is possible to deliver \gls{IDA} results for a complete shot on a between-discharge relevant time scale (e.g. minutes).

The usage of \gls{IMAS} for input and output of the \gls{IDA} framework and the tools around it, such as the database creator and the machine learning tools, make the framework quite flexible to deploy on any machine that can deliver \gls{IMAS}-compliant machine description files and measurement signals. 

Finally, we want to emphasize that the results shown in section \ref{sect:results} should not be interpreted as a quantitative study of the expected uncertainties of plasma profiles and the equilibrium in \Gls{ITER}. While our code should be capable of supporting such a study after more synthetic diagnostics have been added, more work on high-fidelity synthetic diagnostic models is needed in order to consider realistic systematic uncertainties in the analysis (see section \ref{sect:art_data}).

To summarize, a new \gls{IDA} framework has been developed that allows for the simultaneous reconstruction of the magnetic equilibrium and the kinetic profiles. The efficacy of the code was showcased using an \Gls{ITER}-like full-field DT scenario. The combined \gls{MAP} analysis of magnetic measurements, \gls{TS}, interferometry, and polarimetry with realistic diagnostic geometry could be completed under \SI{3}{\minute} on a single server node, including uncertainty propagation. The results of the \gls{MAP} inference were verified with \gls{MCMC}, and good agreement in the uncertainties was found for quantities that were well constrained by measurements.
 
\begin{acknowledgments}
    Work supported using General Atomics Corporate Funds. Thanks for S. Kruger for his suggestions that helped to improve this paper.
\end{acknowledgments}

\bibliography{EQ_prof}% Produces the bibliography via BibTeX.

\begin{thebibliography}{41}%
\makeatletter
\providecommand \@ifxundefined [1]{%
 \@ifx{#1\undefined}
}%
\providecommand \@ifnum [1]{%
 \ifnum #1\expandafter \@firstoftwo
 \else \expandafter \@secondoftwo
 \fi
}%
\providecommand \@ifx [1]{%
 \ifx #1\expandafter \@firstoftwo
 \else \expandafter \@secondoftwo
 \fi
}%
\providecommand \natexlab [1]{#1}%
\providecommand \enquote  [1]{``#1''}%
\providecommand \bibnamefont  [1]{#1}%
\providecommand \bibfnamefont [1]{#1}%
\providecommand \citenamefont [1]{#1}%
\providecommand \href@noop [0]{\@secondoftwo}%
\providecommand \href [0]{\begingroup \@sanitize@url \@href}%
\providecommand \@href[1]{\@@startlink{#1}\@@href}%
\providecommand \@@href[1]{\endgroup#1\@@endlink}%
\providecommand \@sanitize@url [0]{\catcode `\\12\catcode `\$12\catcode
  `\&12\catcode `\#12\catcode `\^12\catcode `\_12\catcode `\%12\relax}%
\providecommand \@@startlink[1]{}%
\providecommand \@@endlink[0]{}%
\providecommand \url  [0]{\begingroup\@sanitize@url \@url }%
\providecommand \@url [1]{\endgroup\@href {#1}{\urlprefix }}%
\providecommand \urlprefix  [0]{URL }%
\providecommand \Eprint [0]{\href }%
\providecommand \doibase [0]{http://dx.doi.org/}%
\providecommand \selectlanguage [0]{\@gobble}%
\providecommand \bibinfo  [0]{\@secondoftwo}%
\providecommand \bibfield  [0]{\@secondoftwo}%
\providecommand \translation [1]{[#1]}%
\providecommand \BibitemOpen [0]{}%
\providecommand \bibitemStop [0]{}%
\providecommand \bibitemNoStop [0]{.\EOS\space}%
\providecommand \EOS [0]{\spacefactor3000\relax}%
\providecommand \BibitemShut  [1]{\csname bibitem#1\endcsname}%
\let\auto@bib@innerbib\@empty
%</preamble>
\bibitem [{\citenamefont {Zvonkov}\ \emph {et~al.}(2016)\citenamefont
  {Zvonkov}, \citenamefont {De~Bock}, \citenamefont {Serov},\ and\
  \citenamefont {Tugarinov}}]{zvonkov2016cxrs}%
  \BibitemOpen
  \bibfield  {author} {\bibinfo {author} {\bibfnamefont {A.}~\bibnamefont
  {Zvonkov}}, \bibinfo {author} {\bibfnamefont {M.}~\bibnamefont {De~Bock}},
  \bibinfo {author} {\bibfnamefont {V.}~\bibnamefont {Serov}}, \ and\ \bibinfo
  {author} {\bibfnamefont {S.}~\bibnamefont {Tugarinov}},\ }\bibfield  {title}
  {\enquote {\bibinfo {title} {{CXRS}-edge {Diagnostic} in the {Harsh} {ITER}
  {Environment}},}\ }in\ \href@noop {} {\emph {\bibinfo {booktitle} {FIP/P4-17,
  26th Fusion Energy Conference (FEC)}}}\ (\bibinfo {year} {2016})\ pp.\
  \bibinfo {pages} {17--22}\BibitemShut {NoStop}%
\bibitem [{\citenamefont {Litnovsky}\ \emph {et~al.}(2015)\citenamefont
  {Litnovsky}, \citenamefont {Matveeva}, \citenamefont {Buzi}, \citenamefont
  {Vera}, \citenamefont {Krasikov}, \citenamefont {Kotov}, \citenamefont
  {Panin}, \citenamefont {Wienhold}, \citenamefont {Philipps}, \citenamefont
  {Bardawil} \emph {et~al.}}]{litnovsky2015studies}%
  \BibitemOpen
  \bibfield  {author} {\bibinfo {author} {\bibfnamefont {A.}~\bibnamefont
  {Litnovsky}}, \bibinfo {author} {\bibfnamefont {M.}~\bibnamefont {Matveeva}},
  \bibinfo {author} {\bibfnamefont {L.}~\bibnamefont {Buzi}}, \bibinfo {author}
  {\bibfnamefont {L.}~\bibnamefont {Vera}}, \bibinfo {author} {\bibfnamefont
  {Y.}~\bibnamefont {Krasikov}}, \bibinfo {author} {\bibfnamefont
  {V.}~\bibnamefont {Kotov}}, \bibinfo {author} {\bibfnamefont
  {A.}~\bibnamefont {Panin}}, \bibinfo {author} {\bibfnamefont
  {P.}~\bibnamefont {Wienhold}}, \bibinfo {author} {\bibfnamefont
  {V.}~\bibnamefont {Philipps}}, \bibinfo {author} {\bibfnamefont {D.~C.}\
  \bibnamefont {Bardawil}},  \emph {et~al.},\ }\bibfield  {title} {\enquote
  {\bibinfo {title} {Studies of protection and recovery techniques of
  diagnostic mirrors for {ITER}},}\ }\href@noop {} {\bibfield  {journal}
  {\bibinfo  {journal} {Nuclear Fusion}\ }\textbf {\bibinfo {volume} {55}},\
  \bibinfo {pages} {093015} (\bibinfo {year} {2015})}\BibitemShut {NoStop}%
\bibitem [{\citenamefont {Walsh}\ \emph {et~al.}(2011)\citenamefont {Walsh},
  \citenamefont {Andrew}, \citenamefont {Barnsley}, \citenamefont {Bertalot},
  \citenamefont {Boivin}, \citenamefont {Bora}, \citenamefont {Bouhamou},
  \citenamefont {Ciattaglia}, \citenamefont {Costley}, \citenamefont {Counsell}
  \emph {et~al.}}]{walsh2011iter}%
  \BibitemOpen
  \bibfield  {author} {\bibinfo {author} {\bibfnamefont {M.}~\bibnamefont
  {Walsh}}, \bibinfo {author} {\bibfnamefont {P.}~\bibnamefont {Andrew}},
  \bibinfo {author} {\bibfnamefont {R.}~\bibnamefont {Barnsley}}, \bibinfo
  {author} {\bibfnamefont {L.}~\bibnamefont {Bertalot}}, \bibinfo {author}
  {\bibfnamefont {R.}~\bibnamefont {Boivin}}, \bibinfo {author} {\bibfnamefont
  {D.}~\bibnamefont {Bora}}, \bibinfo {author} {\bibfnamefont {R.}~\bibnamefont
  {Bouhamou}}, \bibinfo {author} {\bibfnamefont {S.}~\bibnamefont
  {Ciattaglia}}, \bibinfo {author} {\bibfnamefont {A.}~\bibnamefont {Costley}},
  \bibinfo {author} {\bibfnamefont {G.}~\bibnamefont {Counsell}},  \emph
  {et~al.},\ }\bibfield  {title} {\enquote {\bibinfo {title} {{ITER} diagnostic
  challenges},}\ }in\ \href@noop {} {\emph {\bibinfo {booktitle} {2011
  IEEE/NPSS 24th Symposium on Fusion Engineering}}}\ (\bibinfo {organization}
  {IEEE},\ \bibinfo {year} {2011})\ pp.\ \bibinfo {pages} {1--8}\BibitemShut
  {NoStop}%
\bibitem [{\citenamefont {Costley}\ \emph {et~al.}(2005)\citenamefont
  {Costley}, \citenamefont {Sugie}, \citenamefont {Vayakis},\ and\
  \citenamefont {Walker}}]{costley2005technological}%
  \BibitemOpen
  \bibfield  {author} {\bibinfo {author} {\bibfnamefont {A.}~\bibnamefont
  {Costley}}, \bibinfo {author} {\bibfnamefont {T.}~\bibnamefont {Sugie}},
  \bibinfo {author} {\bibfnamefont {G.}~\bibnamefont {Vayakis}}, \ and\
  \bibinfo {author} {\bibfnamefont {C.}~\bibnamefont {Walker}},\ }\bibfield
  {title} {\enquote {\bibinfo {title} {Technological challenges of {ITER}
  diagnostics},}\ }\href@noop {} {\bibfield  {journal} {\bibinfo  {journal}
  {Fusion Engineering and Design}\ }\textbf {\bibinfo {volume} {74}},\ \bibinfo
  {pages} {109--119} (\bibinfo {year} {2005})}\BibitemShut {NoStop}%
\bibitem [{\citenamefont {Fischer}\ \emph {et~al.}(2010)\citenamefont
  {Fischer}, \citenamefont {Fuchs}, \citenamefont {Kurzan}, \citenamefont
  {Suttrop}, \citenamefont {Wolfrum},\ and\ \citenamefont {{ASDEX Upgrade
  Team}}}]{fischer2010integrated}%
  \BibitemOpen
  \bibfield  {author} {\bibinfo {author} {\bibfnamefont {R.}~\bibnamefont
  {Fischer}}, \bibinfo {author} {\bibfnamefont {C.}~\bibnamefont {Fuchs}},
  \bibinfo {author} {\bibfnamefont {B.}~\bibnamefont {Kurzan}}, \bibinfo
  {author} {\bibfnamefont {W.}~\bibnamefont {Suttrop}}, \bibinfo {author}
  {\bibfnamefont {E.}~\bibnamefont {Wolfrum}}, \ and\ \bibinfo {author}
  {\bibnamefont {{ASDEX Upgrade Team}}},\ }\bibfield  {title} {\enquote
  {\bibinfo {title} {Integrated data analysis of profile diagnostics at {ASDEX}
  {Upgrade}},}\ }\href@noop {} {\bibfield  {journal} {\bibinfo  {journal}
  {Fusion science and technology}\ }\textbf {\bibinfo {volume} {58}},\ \bibinfo
  {pages} {675--684} (\bibinfo {year} {2010})}\BibitemShut {NoStop}%
\bibitem [{\citenamefont {Verdoolaege}\ \emph {et~al.}(2010)\citenamefont
  {Verdoolaege}, \citenamefont {Fischer}, \citenamefont {Van~Oost},
  \citenamefont {Contributors} \emph {et~al.}}]{verdoolaege2010potential}%
  \BibitemOpen
  \bibfield  {author} {\bibinfo {author} {\bibfnamefont {G.}~\bibnamefont
  {Verdoolaege}}, \bibinfo {author} {\bibfnamefont {R.}~\bibnamefont
  {Fischer}}, \bibinfo {author} {\bibfnamefont {G.}~\bibnamefont {Van~Oost}},
  \bibinfo {author} {\bibfnamefont {J.-E.}\ \bibnamefont {Contributors}},
  \emph {et~al.},\ }\bibfield  {title} {\enquote {\bibinfo {title} {Potential
  of a {Bayesian} {Integrated} {Determination} of the {Ion} {Effective}
  {Charge} via {Bremsstrahlung} and {Charge} {Exchange} {Spectroscopy} in
  {Tokamak} {Plasmas}},}\ }\href@noop {} {\bibfield  {journal} {\bibinfo
  {journal} {IEEE transactions on plasma science}\ }\textbf {\bibinfo {volume}
  {38}},\ \bibinfo {pages} {3168--3196} (\bibinfo {year} {2010})}\BibitemShut
  {NoStop}%
\bibitem [{\citenamefont {Fischer}\ \emph {et~al.}(2024)\citenamefont
  {Fischer}, \citenamefont {Bock}, \citenamefont {Denk}, \citenamefont
  {Salewski}, \citenamefont {Schneider}, \citenamefont {Stieglitz},\ and\
  \citenamefont {{ASDEX Upgrade
  Team}}}]{fischer2024integrateddataanalysisvalidation}%
  \BibitemOpen
  \bibfield  {author} {\bibinfo {author} {\bibfnamefont {R.}~\bibnamefont
  {Fischer}}, \bibinfo {author} {\bibfnamefont {A.}~\bibnamefont {Bock}},
  \bibinfo {author} {\bibfnamefont {S.~S.}\ \bibnamefont {Denk}}, \bibinfo
  {author} {\bibfnamefont {A.~M.~M.}\ \bibnamefont {Salewski}}, \bibinfo
  {author} {\bibfnamefont {M.}~\bibnamefont {Schneider}}, \bibinfo {author}
  {\bibfnamefont {D.}~\bibnamefont {Stieglitz}}, \ and\ \bibinfo {author}
  {\bibnamefont {{ASDEX Upgrade Team}}},\ }\href
  {https://arxiv.org/abs/2411.09270} {\enquote {\bibinfo {title} {{Integrated}
  {Data} {Analysis} and {Validation}},}\ } (\bibinfo {year} {2024}),\ \Eprint
  {http://arxiv.org/abs/2411.09270} {arXiv:2411.09270 [physics.plasm-ph]}
  \BibitemShut {NoStop}%
\bibitem [{\citenamefont {Pavone}\ \emph {et~al.}(2023)\citenamefont {Pavone},
  \citenamefont {Merlo}, \citenamefont {Kwak},\ and\ \citenamefont
  {Svensson}}]{Pavone_2023}%
  \BibitemOpen
  \bibfield  {author} {\bibinfo {author} {\bibfnamefont {A.}~\bibnamefont
  {Pavone}}, \bibinfo {author} {\bibfnamefont {A.}~\bibnamefont {Merlo}},
  \bibinfo {author} {\bibfnamefont {S.}~\bibnamefont {Kwak}}, \ and\ \bibinfo
  {author} {\bibfnamefont {J.}~\bibnamefont {Svensson}},\ }\bibfield  {title}
  {\enquote {\bibinfo {title} {Machine learning and {Bayesian} inference in
  nuclear fusion research: an overview},}\ }\href {\doibase
  10.1088/1361-6587/acc60f} {\bibfield  {journal} {\bibinfo  {journal} {Plasma
  Physics and Controlled Fusion}\ }\textbf {\bibinfo {volume} {65}},\ \bibinfo
  {pages} {053001} (\bibinfo {year} {2023})}\BibitemShut {NoStop}%
\bibitem [{\citenamefont {Lao}\ \emph {et~al.}(1985)\citenamefont {Lao},
  \citenamefont {John}, \citenamefont {Stambaugh}, \citenamefont {Kellman},\
  and\ \citenamefont {Pfeiffer}}]{lao1985efit}%
  \BibitemOpen
  \bibfield  {author} {\bibinfo {author} {\bibfnamefont {L.~L.}\ \bibnamefont
  {Lao}}, \bibinfo {author} {\bibfnamefont {H.~S.}\ \bibnamefont {John}},
  \bibinfo {author} {\bibfnamefont {R.}~\bibnamefont {Stambaugh}}, \bibinfo
  {author} {\bibfnamefont {A.}~\bibnamefont {Kellman}}, \ and\ \bibinfo
  {author} {\bibfnamefont {W.}~\bibnamefont {Pfeiffer}},\ }\bibfield  {title}
  {\enquote {\bibinfo {title} {Reconstruction of current profile parameters and
  plasma shapes in tokamaks},}\ }\href {\doibase 10.1088/0029-5515/25/11/007}
  {\bibfield  {journal} {\bibinfo  {journal} {Nuclear Fusion}\ }\textbf
  {\bibinfo {volume} {25}},\ \bibinfo {pages} {1611} (\bibinfo {year}
  {1985})}\BibitemShut {NoStop}%
\bibitem [{\citenamefont {Lao}\ \emph {et~al.}(2005)\citenamefont {Lao},
  \citenamefont {John}, \citenamefont {Peng}, \citenamefont {Ferron},
  \citenamefont {Strait}, \citenamefont {Taylor}, \citenamefont {Meyer},
  \citenamefont {Zhang},\ and\ \citenamefont {You}}]{lao2005efit}%
  \BibitemOpen
  \bibfield  {author} {\bibinfo {author} {\bibfnamefont {L.~L.}\ \bibnamefont
  {Lao}}, \bibinfo {author} {\bibfnamefont {H.~E.~S.}\ \bibnamefont {John}},
  \bibinfo {author} {\bibfnamefont {Q.}~\bibnamefont {Peng}}, \bibinfo {author}
  {\bibfnamefont {J.~R.}\ \bibnamefont {Ferron}}, \bibinfo {author}
  {\bibfnamefont {E.~J.}\ \bibnamefont {Strait}}, \bibinfo {author}
  {\bibfnamefont {T.~S.}\ \bibnamefont {Taylor}}, \bibinfo {author}
  {\bibfnamefont {W.~H.}\ \bibnamefont {Meyer}}, \bibinfo {author}
  {\bibfnamefont {C.}~\bibnamefont {Zhang}}, \ and\ \bibinfo {author}
  {\bibfnamefont {K.~I.}\ \bibnamefont {You}},\ }\bibfield  {title} {\enquote
  {\bibinfo {title} {{MHD} {Equilibrium} {Reconstruction} in the {DIII-D}
  {Tokamak}},}\ }\href {\doibase 10.13182/FST48-968} {\bibfield  {journal}
  {\bibinfo  {journal} {Fusion Science and Technology}\ }\textbf {\bibinfo
  {volume} {48}},\ \bibinfo {pages} {968--977} (\bibinfo {year} {2005})},\
  \Eprint {http://arxiv.org/abs/https://doi.org/10.13182/FST48-968}
  {https://doi.org/10.13182/FST48-968} \BibitemShut {NoStop}%
\bibitem [{\citenamefont {Fischer}, \citenamefont {Dinklage},\ and\
  \citenamefont {Pasch}(2003)}]{fischer2003bayesian}%
  \BibitemOpen
  \bibfield  {author} {\bibinfo {author} {\bibfnamefont {R.}~\bibnamefont
  {Fischer}}, \bibinfo {author} {\bibfnamefont {A.}~\bibnamefont {Dinklage}}, \
  and\ \bibinfo {author} {\bibfnamefont {E.}~\bibnamefont {Pasch}},\ }\bibfield
   {title} {\enquote {\bibinfo {title} {Bayesian modelling of fusion
  diagnostics},}\ }\href@noop {} {\bibfield  {journal} {\bibinfo  {journal}
  {Plasma Physics and Controlled Fusion}\ }\textbf {\bibinfo {volume} {45}},\
  \bibinfo {pages} {1095} (\bibinfo {year} {2003})}\BibitemShut {NoStop}%
\bibitem [{\citenamefont {Avdeeva}\ \emph {et~al.}(2024)\citenamefont
  {Avdeeva}, \citenamefont {Thome}, \citenamefont {Berkery}, \citenamefont
  {Kaye}, \citenamefont {McClenaghan}, \citenamefont {Meneghini}, \citenamefont
  {Odstrcil}, \citenamefont {Sabbagh}, \citenamefont {Smith},\ and\
  \citenamefont {Turnbull}}]{avdeeva2024accuracy}%
  \BibitemOpen
  \bibfield  {author} {\bibinfo {author} {\bibfnamefont {G.}~\bibnamefont
  {Avdeeva}}, \bibinfo {author} {\bibfnamefont {K.~E.}\ \bibnamefont {Thome}},
  \bibinfo {author} {\bibfnamefont {J.~W.}\ \bibnamefont {Berkery}}, \bibinfo
  {author} {\bibfnamefont {S.~M.}\ \bibnamefont {Kaye}}, \bibinfo {author}
  {\bibfnamefont {J.}~\bibnamefont {McClenaghan}}, \bibinfo {author}
  {\bibfnamefont {O.}~\bibnamefont {Meneghini}}, \bibinfo {author}
  {\bibfnamefont {T.}~\bibnamefont {Odstrcil}}, \bibinfo {author}
  {\bibfnamefont {S.~A.}\ \bibnamefont {Sabbagh}}, \bibinfo {author}
  {\bibfnamefont {S.~P.}\ \bibnamefont {Smith}}, \ and\ \bibinfo {author}
  {\bibfnamefont {A.~D.}\ \bibnamefont {Turnbull}},\ }\bibfield  {title}
  {\enquote {\bibinfo {title} {Accuracy of kinetic equilibrium reconstruction
  of {NSTX} and {NSTX-U} plasmas and its impact on the transport and stability
  analysis},}\ }\href@noop {} {\bibfield  {journal} {\bibinfo  {journal}
  {Plasma Physics and Controlled Fusion}\ }\textbf {\bibinfo {volume} {66}},\
  \bibinfo {pages} {115003} (\bibinfo {year} {2024})}\BibitemShut {NoStop}%
\bibitem [{\citenamefont {Svensson}\ \emph {et~al.}(2004)\citenamefont
  {Svensson}, \citenamefont {Dinklage}, \citenamefont {Geiger}, \citenamefont
  {Werner},\ and\ \citenamefont {Fischer}}]{svensson2004integrating}%
  \BibitemOpen
  \bibfield  {author} {\bibinfo {author} {\bibfnamefont {J.}~\bibnamefont
  {Svensson}}, \bibinfo {author} {\bibfnamefont {A.}~\bibnamefont {Dinklage}},
  \bibinfo {author} {\bibfnamefont {J.}~\bibnamefont {Geiger}}, \bibinfo
  {author} {\bibfnamefont {A.}~\bibnamefont {Werner}}, \ and\ \bibinfo {author}
  {\bibfnamefont {R.}~\bibnamefont {Fischer}},\ }\bibfield  {title} {\enquote
  {\bibinfo {title} {Integrating diagnostic data analysis for {W7-AS} using
  {Bayesian} graphical models},}\ }\href@noop {} {\bibfield  {journal}
  {\bibinfo  {journal} {Review of Scientific Instruments}\ }\textbf {\bibinfo
  {volume} {75}},\ \bibinfo {pages} {4219--4221} (\bibinfo {year}
  {2004})}\BibitemShut {NoStop}%
\bibitem [{\citenamefont {Svensson}\ and\ \citenamefont
  {Werner}(2007)}]{svensson2007large}%
  \BibitemOpen
  \bibfield  {author} {\bibinfo {author} {\bibfnamefont {J.}~\bibnamefont
  {Svensson}}\ and\ \bibinfo {author} {\bibfnamefont {A.}~\bibnamefont
  {Werner}},\ }\bibfield  {title} {\enquote {\bibinfo {title} {Large scale
  {Bayesian} data analysis for nuclear fusion experiments},}\ }in\ \href@noop
  {} {\emph {\bibinfo {booktitle} {2007 IEEE International Symposium on
  Intelligent Signal Processing}}}\ (\bibinfo {organization} {IEEE},\ \bibinfo
  {year} {2007})\ pp.\ \bibinfo {pages} {1--6}\BibitemShut {NoStop}%
\bibitem [{\citenamefont {Kruger}\ \emph {et~al.}(2024)\citenamefont {Kruger},
  \citenamefont {Leddy}, \citenamefont {Howell}, \citenamefont {Madireddy},
  \citenamefont {Akcay}, \citenamefont {Bechtel~Amara}, \citenamefont
  {McClenaghan}, \citenamefont {Lao}, \citenamefont {Orozco}, \citenamefont
  {Smith} \emph {et~al.}}]{kruger2024thinking}%
  \BibitemOpen
  \bibfield  {author} {\bibinfo {author} {\bibfnamefont {S.}~\bibnamefont
  {Kruger}}, \bibinfo {author} {\bibfnamefont {J.}~\bibnamefont {Leddy}},
  \bibinfo {author} {\bibfnamefont {E.}~\bibnamefont {Howell}}, \bibinfo
  {author} {\bibfnamefont {S.}~\bibnamefont {Madireddy}}, \bibinfo {author}
  {\bibfnamefont {C.}~\bibnamefont {Akcay}}, \bibinfo {author} {\bibfnamefont
  {T.}~\bibnamefont {Bechtel~Amara}}, \bibinfo {author} {\bibfnamefont
  {J.}~\bibnamefont {McClenaghan}}, \bibinfo {author} {\bibfnamefont
  {L.}~\bibnamefont {Lao}}, \bibinfo {author} {\bibfnamefont {D.}~\bibnamefont
  {Orozco}}, \bibinfo {author} {\bibfnamefont {S.}~\bibnamefont {Smith}},
  \emph {et~al.},\ }\bibfield  {title} {\enquote {\bibinfo {title} {Thinking
  {Bayesian} for plasma physicists},}\ }\href@noop {} {\bibfield  {journal}
  {\bibinfo  {journal} {Physics of Plasmas}\ }\textbf {\bibinfo {volume} {31}}
  (\bibinfo {year} {2024})}\BibitemShut {NoStop}%
\bibitem [{\citenamefont {Bergmann}\ \emph {et~al.}(2024)\citenamefont
  {Bergmann}, \citenamefont {Fischer}, \citenamefont {Angioni}, \citenamefont
  {H{\"o}fler}, \citenamefont {Cabrera}, \citenamefont {G{\"o}rler},
  \citenamefont {Luda}, \citenamefont {Bilato}, \citenamefont {Tardini},
  \citenamefont {Jenko} \emph {et~al.}}]{bergmann2024plasma}%
  \BibitemOpen
  \bibfield  {author} {\bibinfo {author} {\bibfnamefont {M.}~\bibnamefont
  {Bergmann}}, \bibinfo {author} {\bibfnamefont {R.}~\bibnamefont {Fischer}},
  \bibinfo {author} {\bibfnamefont {C.}~\bibnamefont {Angioni}}, \bibinfo
  {author} {\bibfnamefont {K.}~\bibnamefont {H{\"o}fler}}, \bibinfo {author}
  {\bibfnamefont {P.~M.}\ \bibnamefont {Cabrera}}, \bibinfo {author}
  {\bibfnamefont {T.}~\bibnamefont {G{\"o}rler}}, \bibinfo {author}
  {\bibfnamefont {T.}~\bibnamefont {Luda}}, \bibinfo {author} {\bibfnamefont
  {R.}~\bibnamefont {Bilato}}, \bibinfo {author} {\bibfnamefont
  {G.}~\bibnamefont {Tardini}}, \bibinfo {author} {\bibfnamefont
  {F.}~\bibnamefont {Jenko}},  \emph {et~al.},\ }\bibfield  {title} {\enquote
  {\bibinfo {title} {Plasma profile reconstruction supported by kinetic
  modeling},}\ }\href@noop {} {\bibfield  {journal} {\bibinfo  {journal}
  {Nuclear Fusion}\ }\textbf {\bibinfo {volume} {64}},\ \bibinfo {pages}
  {056024} (\bibinfo {year} {2024})}\BibitemShut {NoStop}%
\bibitem [{\citenamefont {Nishizawa}\ \emph
  {et~al.}(2022{\natexlab{a}})\citenamefont {Nishizawa}, \citenamefont {Dux},
  \citenamefont {McDermott}, \citenamefont {Sciortino}, \citenamefont
  {Cavedon}, \citenamefont {Schuster}, \citenamefont {Wolfrum}, \citenamefont
  {Von~Toussaint}, \citenamefont {Van~Vuuren}, \citenamefont {Cruz-Zabala}
  \emph {et~al.}}]{nishizawa2022non}%
  \BibitemOpen
  \bibfield  {author} {\bibinfo {author} {\bibfnamefont {T.}~\bibnamefont
  {Nishizawa}}, \bibinfo {author} {\bibfnamefont {R.}~\bibnamefont {Dux}},
  \bibinfo {author} {\bibfnamefont {R.}~\bibnamefont {McDermott}}, \bibinfo
  {author} {\bibfnamefont {F.}~\bibnamefont {Sciortino}}, \bibinfo {author}
  {\bibfnamefont {M.}~\bibnamefont {Cavedon}}, \bibinfo {author} {\bibfnamefont
  {C.}~\bibnamefont {Schuster}}, \bibinfo {author} {\bibfnamefont
  {E.}~\bibnamefont {Wolfrum}}, \bibinfo {author} {\bibfnamefont
  {U.}~\bibnamefont {Von~Toussaint}}, \bibinfo {author} {\bibfnamefont {A.~J.}\
  \bibnamefont {Van~Vuuren}}, \bibinfo {author} {\bibfnamefont {D.~J.}\
  \bibnamefont {Cruz-Zabala}},  \emph {et~al.},\ }\bibfield  {title} {\enquote
  {\bibinfo {title} {Non-parametric inference of impurity transport
  coefficients in the {ASDEX} {Upgrade} tokamak},}\ }\href@noop {} {\bibfield
  {journal} {\bibinfo  {journal} {Nuclear Fusion}\ }\textbf {\bibinfo {volume}
  {62}},\ \bibinfo {pages} {076021} (\bibinfo {year}
  {2022}{\natexlab{a}})}\BibitemShut {NoStop}%
\bibitem [{\citenamefont {Fischer}\ \emph {et~al.}(2016)\citenamefont
  {Fischer}, \citenamefont {Bock}, \citenamefont {Dunne}, \citenamefont
  {Fuchs}, \citenamefont {Giannone}, \citenamefont {Lackner}, \citenamefont
  {McCarthy}, \citenamefont {Poli}, \citenamefont {Preuss}, \citenamefont
  {Rampp}, \citenamefont {Schubert}, \citenamefont {Stober}, \citenamefont
  {Suttrop}, \citenamefont {Tardini}, \citenamefont {Weiland},\ and\
  \citenamefont {{ASDEX Upgrade Team}}}]{IDE}%
  \BibitemOpen
  \bibfield  {author} {\bibinfo {author} {\bibfnamefont {R.}~\bibnamefont
  {Fischer}}, \bibinfo {author} {\bibfnamefont {A.}~\bibnamefont {Bock}},
  \bibinfo {author} {\bibfnamefont {M.}~\bibnamefont {Dunne}}, \bibinfo
  {author} {\bibfnamefont {J.~C.}\ \bibnamefont {Fuchs}}, \bibinfo {author}
  {\bibfnamefont {L.}~\bibnamefont {Giannone}}, \bibinfo {author}
  {\bibfnamefont {K.}~\bibnamefont {Lackner}}, \bibinfo {author} {\bibfnamefont
  {P.~J.}\ \bibnamefont {McCarthy}}, \bibinfo {author} {\bibfnamefont
  {E.}~\bibnamefont {Poli}}, \bibinfo {author} {\bibfnamefont {R.}~\bibnamefont
  {Preuss}}, \bibinfo {author} {\bibfnamefont {M.}~\bibnamefont {Rampp}},
  \bibinfo {author} {\bibfnamefont {M.}~\bibnamefont {Schubert}}, \bibinfo
  {author} {\bibfnamefont {J.}~\bibnamefont {Stober}}, \bibinfo {author}
  {\bibfnamefont {W.}~\bibnamefont {Suttrop}}, \bibinfo {author} {\bibfnamefont
  {G.}~\bibnamefont {Tardini}}, \bibinfo {author} {\bibfnamefont
  {M.}~\bibnamefont {Weiland}}, \ and\ \bibinfo {author} {\bibnamefont {{ASDEX
  Upgrade Team}}},\ }\bibfield  {title} {\enquote {\bibinfo {title} {Coupling
  of the {Flux} {Diffusion} {Equation} with the {Equilibrium} {Reconstruction}
  at {ASDEX Upgrade}},}\ }\href {\doibase 10.13182/FST15-185} {\bibfield
  {journal} {\bibinfo  {journal} {Fusion Science and Technology}\ }\textbf
  {\bibinfo {volume} {69}},\ \bibinfo {pages} {526--536} (\bibinfo {year}
  {2016})},\ \Eprint {http://arxiv.org/abs/https://doi.org/10.13182/FST15-185}
  {https://doi.org/10.13182/FST15-185} \BibitemShut {NoStop}%
\bibitem [{\citenamefont {Smith}\ \emph {et~al.}(2024)\citenamefont {Smith},
  \citenamefont {Xing}, \citenamefont {Amara}, \citenamefont {Denk},
  \citenamefont {DeShazer} \emph {et~al.}}]{smith2024xloop}%
  \BibitemOpen
  \bibfield  {author} {\bibinfo {author} {\bibfnamefont {S.}~\bibnamefont
  {Smith}}, \bibinfo {author} {\bibfnamefont {Z.~A.}\ \bibnamefont {Xing}},
  \bibinfo {author} {\bibfnamefont {T.~B.}\ \bibnamefont {Amara}}, \bibinfo
  {author} {\bibfnamefont {S.~S.}\ \bibnamefont {Denk}}, \bibinfo {author}
  {\bibfnamefont {E.~W.}\ \bibnamefont {DeShazer}},  \emph {et~al.},\
  }\bibfield  {title} {\enquote {\bibinfo {title} {Expediting {Higher}
  {Fidelity} {Plasma} {State} {Reconstructions} for the {DIII-D} {National}
  {Fusion} {Facility} {Using} {Leadership} {Class} {Computing} {Resources}},}\
  }in\ \href {\doibase 10.1109/SCW63240.2024.00265} {\emph {\bibinfo
  {booktitle} {SC24-W: Workshops of the International Conference for High
  Performance Computing, Networking, Storage and Analysis}}}\ (\bibinfo
  {publisher} {Association of Computing Machinery Digital Library},\ \bibinfo
  {year} {2024})\ pp.\ \bibinfo {pages} {2118--2126}\BibitemShut {NoStop}%
\bibitem [{\citenamefont {Rasmussen}(2003)}]{rasmussen2003gaussian}%
  \BibitemOpen
  \bibfield  {author} {\bibinfo {author} {\bibfnamefont {C.~E.}\ \bibnamefont
  {Rasmussen}},\ }\bibfield  {title} {\enquote {\bibinfo {title} {Gaussian
  processes in machine learning},}\ }in\ \href@noop {} {\emph {\bibinfo
  {booktitle} {Summer school on machine learning}}}\ (\bibinfo  {publisher}
  {Springer},\ \bibinfo {year} {2003})\ pp.\ \bibinfo {pages}
  {63--71}\BibitemShut {NoStop}%
\bibitem [{\citenamefont {Kwak}\ \emph {et~al.}(2022)\citenamefont {Kwak},
  \citenamefont {Svensson}, \citenamefont {Ford}, \citenamefont {Appel},
  \citenamefont {Ghim},\ and\ \citenamefont {Contributors}}]{kwak2022bayesian}%
  \BibitemOpen
  \bibfield  {author} {\bibinfo {author} {\bibfnamefont {S.}~\bibnamefont
  {Kwak}}, \bibinfo {author} {\bibfnamefont {J.}~\bibnamefont {Svensson}},
  \bibinfo {author} {\bibfnamefont {O.}~\bibnamefont {Ford}}, \bibinfo {author}
  {\bibfnamefont {L.}~\bibnamefont {Appel}}, \bibinfo {author} {\bibfnamefont
  {Y.-c.}\ \bibnamefont {Ghim}}, \ and\ \bibinfo {author} {\bibfnamefont
  {J.}~\bibnamefont {Contributors}},\ }\bibfield  {title} {\enquote {\bibinfo
  {title} {Bayesian inference of axisymmetric plasma equilibrium},}\
  }\href@noop {} {\bibfield  {journal} {\bibinfo  {journal} {Nuclear Fusion}\
  }\textbf {\bibinfo {volume} {62}},\ \bibinfo {pages} {126069} (\bibinfo
  {year} {2022})}\BibitemShut {NoStop}%
\bibitem [{\citenamefont {Merlo}\ \emph {et~al.}(2023)\citenamefont {Merlo},
  \citenamefont {Pavone}, \citenamefont {B{\"o}ckenhoff}, \citenamefont
  {Pasch}, \citenamefont {Fuchert}, \citenamefont {Brunner}, \citenamefont
  {Rahbarnia}, \citenamefont {Schilling}, \citenamefont {H{\"o}fel},
  \citenamefont {Kwak} \emph {et~al.}}]{merlo2023accelerated}%
  \BibitemOpen
  \bibfield  {author} {\bibinfo {author} {\bibfnamefont {A.}~\bibnamefont
  {Merlo}}, \bibinfo {author} {\bibfnamefont {A.}~\bibnamefont {Pavone}},
  \bibinfo {author} {\bibfnamefont {D.}~\bibnamefont {B{\"o}ckenhoff}},
  \bibinfo {author} {\bibfnamefont {E.}~\bibnamefont {Pasch}}, \bibinfo
  {author} {\bibfnamefont {G.}~\bibnamefont {Fuchert}}, \bibinfo {author}
  {\bibfnamefont {K.~J.}\ \bibnamefont {Brunner}}, \bibinfo {author}
  {\bibfnamefont {K.}~\bibnamefont {Rahbarnia}}, \bibinfo {author}
  {\bibfnamefont {J.}~\bibnamefont {Schilling}}, \bibinfo {author}
  {\bibfnamefont {U.}~\bibnamefont {H{\"o}fel}}, \bibinfo {author}
  {\bibfnamefont {S.}~\bibnamefont {Kwak}},  \emph {et~al.},\ }\bibfield
  {title} {\enquote {\bibinfo {title} {Accelerated {Bayesian} inference of
  plasma profiles with self-consistent {MHD} equilibria at {W7-X} via neural
  networks},}\ }\href@noop {} {\bibfield  {journal} {\bibinfo  {journal}
  {Journal of Instrumentation}\ }\textbf {\bibinfo {volume} {18}},\ \bibinfo
  {pages} {P11012} (\bibinfo {year} {2023})}\BibitemShut {NoStop}%
\bibitem [{\citenamefont {de~Vicente}\ \emph {et~al.}(2023)\citenamefont
  {de~Vicente}, \citenamefont {Mazon}, \citenamefont {Xu}, \citenamefont
  {Pinches}, \citenamefont {Churchill}, \citenamefont {Dinklage}, \citenamefont
  {Fischer}, \citenamefont {Murari}, \citenamefont {Rodriguez-Fernandez},
  \citenamefont {Stillerman}, \citenamefont {Vega},\ and\ \citenamefont
  {Verdoolaege}}]{gonzalez2023itpaida}%
  \BibitemOpen
  \bibfield  {author} {\bibinfo {author} {\bibfnamefont {S.~G.}\ \bibnamefont
  {de~Vicente}}, \bibinfo {author} {\bibfnamefont {D.}~\bibnamefont {Mazon}},
  \bibinfo {author} {\bibfnamefont {M.}~\bibnamefont {Xu}}, \bibinfo {author}
  {\bibfnamefont {S.}~\bibnamefont {Pinches}}, \bibinfo {author} {\bibfnamefont
  {M.}~\bibnamefont {Churchill}}, \bibinfo {author} {\bibfnamefont
  {A.}~\bibnamefont {Dinklage}}, \bibinfo {author} {\bibfnamefont
  {R.}~\bibnamefont {Fischer}}, \bibinfo {author} {\bibfnamefont
  {A.}~\bibnamefont {Murari}}, \bibinfo {author} {\bibfnamefont
  {P.}~\bibnamefont {Rodriguez-Fernandez}}, \bibinfo {author} {\bibfnamefont
  {J.}~\bibnamefont {Stillerman}}, \bibinfo {author} {\bibfnamefont
  {J.}~\bibnamefont {Vega}}, \ and\ \bibinfo {author} {\bibfnamefont
  {G.}~\bibnamefont {Verdoolaege}},\ }\bibfield  {title} {\enquote {\bibinfo
  {title} {Summary report of the {4th} {IAEA} {Technical} {Meeting} on {Fusion}
  {Data} {Processing}, {Validation} and {Analysis} {(FDPVA)}},}\ }\href
  {\doibase 10.1088/1741-4326/acbfce} {\bibfield  {journal} {\bibinfo
  {journal} {Nuclear Fusion}\ }\textbf {\bibinfo {volume} {63}},\ \bibinfo
  {pages} {047001} (\bibinfo {year} {2023})}\BibitemShut {NoStop}%
\bibitem [{\citenamefont {Nishizawa}\ \emph
  {et~al.}(2022{\natexlab{b}})\citenamefont {Nishizawa}, \citenamefont {Dux},
  \citenamefont {McDermott}, \citenamefont {Sciortino}, \citenamefont
  {Cavedon}, \citenamefont {Schuster}, \citenamefont {Wolfrum}, \citenamefont
  {von Toussaint}, \citenamefont {Vuuren}, \citenamefont {Cruz-Zabala},
  \citenamefont {Cano-Megias}, \citenamefont {Moon},\ and\ \citenamefont {the
  ASDEX Upgrade~Team}}]{nishizawa2022impurity}%
  \BibitemOpen
  \bibfield  {author} {\bibinfo {author} {\bibfnamefont {T.}~\bibnamefont
  {Nishizawa}}, \bibinfo {author} {\bibfnamefont {R.}~\bibnamefont {Dux}},
  \bibinfo {author} {\bibfnamefont {R.}~\bibnamefont {McDermott}}, \bibinfo
  {author} {\bibfnamefont {F.}~\bibnamefont {Sciortino}}, \bibinfo {author}
  {\bibfnamefont {M.}~\bibnamefont {Cavedon}}, \bibinfo {author} {\bibfnamefont
  {C.}~\bibnamefont {Schuster}}, \bibinfo {author} {\bibfnamefont
  {E.}~\bibnamefont {Wolfrum}}, \bibinfo {author} {\bibfnamefont
  {U.}~\bibnamefont {von Toussaint}}, \bibinfo {author} {\bibfnamefont {A.~V.}\
  \bibnamefont {Vuuren}}, \bibinfo {author} {\bibfnamefont {D.}~\bibnamefont
  {Cruz-Zabala}}, \bibinfo {author} {\bibfnamefont {P.}~\bibnamefont
  {Cano-Megias}}, \bibinfo {author} {\bibfnamefont {C.}~\bibnamefont {Moon}}, \
  and\ \bibinfo {author} {\bibnamefont {the ASDEX Upgrade~Team}},\ }\bibfield
  {title} {\enquote {\bibinfo {title} {Non-parametric inference of impurity
  transport coefficients in the asdex upgrade tokamak},}\ }\href {\doibase
  10.1088/1741-4326/ac60e8} {\bibfield  {journal} {\bibinfo  {journal} {Nuclear
  Fusion}\ }\textbf {\bibinfo {volume} {62}},\ \bibinfo {pages} {076021}
  (\bibinfo {year} {2022}{\natexlab{b}})}\BibitemShut {NoStop}%
\bibitem [{\citenamefont {Fletcher}(2000)}]{fletcher2000practical}%
  \BibitemOpen
  \bibfield  {author} {\bibinfo {author} {\bibfnamefont {R.}~\bibnamefont
  {Fletcher}},\ }\href@noop {} {\emph {\bibinfo {title} {Practical methods of
  optimization}}}\ (\bibinfo  {publisher} {John Wiley \& Sons},\ \bibinfo
  {year} {2000})\BibitemShut {NoStop}%
\bibitem [{\citenamefont {Foreman-Mackey}\ \emph {et~al.}(2013)\citenamefont
  {Foreman-Mackey}, \citenamefont {Hogg}, \citenamefont {Lang},\ and\
  \citenamefont {Goodman}}]{EMCEE}%
  \BibitemOpen
  \bibfield  {author} {\bibinfo {author} {\bibfnamefont {D.}~\bibnamefont
  {Foreman-Mackey}}, \bibinfo {author} {\bibfnamefont {D.~W.}\ \bibnamefont
  {Hogg}}, \bibinfo {author} {\bibfnamefont {D.}~\bibnamefont {Lang}}, \ and\
  \bibinfo {author} {\bibfnamefont {J.}~\bibnamefont {Goodman}},\ }\bibfield
  {title} {\enquote {\bibinfo {title} {emcee: {The} {MCMC} {Hammer}},}\ }\href
  {\doibase 10.1086/670067} {\bibfield  {journal} {\bibinfo  {journal} {arXiv
  preprint arXiv:1202.3665}\ }\textbf {\bibinfo {volume} {125}},\ \bibinfo
  {pages} {306} (\bibinfo {year} {2013})}\BibitemShut {NoStop}%
\bibitem [{\citenamefont {Karamanis}, \citenamefont {Beutler},\ and\
  \citenamefont {Peacock}(2021)}]{karamanis2021zeus}%
  \BibitemOpen
  \bibfield  {author} {\bibinfo {author} {\bibfnamefont {M.}~\bibnamefont
  {Karamanis}}, \bibinfo {author} {\bibfnamefont {F.}~\bibnamefont {Beutler}},
  \ and\ \bibinfo {author} {\bibfnamefont {J.~A.}\ \bibnamefont {Peacock}},\
  }\bibfield  {title} {\enquote {\bibinfo {title} {zeus: A python
  implementation of {Ensemble} {Slice} {Sampling} for efficient {Bayesian}
  parameter inference},}\ }\href@noop {} {\bibfield  {journal} {\bibinfo
  {journal} {arXiv preprint arXiv:2105.03468}\ } (\bibinfo {year}
  {2021})}\BibitemShut {NoStop}%
\bibitem [{\citenamefont {Karamanis}\ and\ \citenamefont
  {Beutler}(2020)}]{karamanis2020ensemble}%
  \BibitemOpen
  \bibfield  {author} {\bibinfo {author} {\bibfnamefont {M.}~\bibnamefont
  {Karamanis}}\ and\ \bibinfo {author} {\bibfnamefont {F.}~\bibnamefont
  {Beutler}},\ }\bibfield  {title} {\enquote {\bibinfo {title} {Ensemble slice
  sampling: {Parallel}, black-box and gradient-free inference for correlated \&
  multimodal distributions},}\ }\href@noop {} {\bibfield  {journal} {\bibinfo
  {journal} {arXiv preprint arXiv: 2002.06212}\ } (\bibinfo {year}
  {2020})}\BibitemShut {NoStop}%
\bibitem [{\citenamefont {Veen}\ and\ \citenamefont
  {Hoekstra}(2020)}]{muscle3}%
  \BibitemOpen
  \bibfield  {author} {\bibinfo {author} {\bibfnamefont {L.~E.}\ \bibnamefont
  {Veen}}\ and\ \bibinfo {author} {\bibfnamefont {A.~G.}\ \bibnamefont
  {Hoekstra}},\ }\bibfield  {title} {\enquote {\bibinfo {title} {Easing
  multiscale model design and coupling with {MUSCLE 3}},}\ }in\ \href@noop {}
  {\emph {\bibinfo {booktitle} {International Conference on Computational
  Science}}}\ (\bibinfo {organization} {Springer},\ \bibinfo {year} {2020})\
  pp.\ \bibinfo {pages} {425--438}\BibitemShut {NoStop}%
\bibitem [{\citenamefont {Denk}\ \emph {et~al.}(2020)\citenamefont {Denk},
  \citenamefont {Fischer}, \citenamefont {Poli}, \citenamefont {Maj},
  \citenamefont {Nielsen}, \citenamefont {Rasmussen}, \citenamefont {Stejner},\
  and\ \citenamefont {Willensdorfer}}]{ECRad}%
  \BibitemOpen
  \bibfield  {author} {\bibinfo {author} {\bibfnamefont {S.}~\bibnamefont
  {Denk}}, \bibinfo {author} {\bibfnamefont {R.}~\bibnamefont {Fischer}},
  \bibinfo {author} {\bibfnamefont {E.}~\bibnamefont {Poli}}, \bibinfo {author}
  {\bibfnamefont {O.}~\bibnamefont {Maj}}, \bibinfo {author} {\bibfnamefont
  {S.}~\bibnamefont {Nielsen}}, \bibinfo {author} {\bibfnamefont
  {J.}~\bibnamefont {Rasmussen}}, \bibinfo {author} {\bibfnamefont
  {M.}~\bibnamefont {Stejner}}, \ and\ \bibinfo {author} {\bibfnamefont
  {M.}~\bibnamefont {Willensdorfer}},\ }\bibfield  {title} {\enquote {\bibinfo
  {title} {{ECRad}: An electron cyclotron radiation transport solver for
  advanced data analysis in thermal and non-thermal fusion plasmas},}\ }\href
  {\doibase https://doi.org/10.1016/j.cpc.2020.107175} {\bibfield  {journal}
  {\bibinfo  {journal} {Computer Physics Communications}\ ,\ \bibinfo {pages}
  {107175}} (\bibinfo {year} {2020})}\BibitemShut {NoStop}%
\bibitem [{\citenamefont {Farina}\ \emph {et~al.}(2008)\citenamefont {Farina},
  \citenamefont {Figini}, \citenamefont {Platania},\ and\ \citenamefont
  {Sozzi}}]{SPECE}%
  \BibitemOpen
  \bibfield  {author} {\bibinfo {author} {\bibfnamefont {D.}~\bibnamefont
  {Farina}}, \bibinfo {author} {\bibfnamefont {L.}~\bibnamefont {Figini}},
  \bibinfo {author} {\bibfnamefont {P.}~\bibnamefont {Platania}}, \ and\
  \bibinfo {author} {\bibfnamefont {C.}~\bibnamefont {Sozzi}},\ }\bibfield
  {title} {\enquote {\bibinfo {title} {{SPECE}: a code for {Electron}
  {Cyclotron} {Emission} in tokamaks},}\ }\href {\doibase 10.1063/1.2905053}
  {\bibfield  {journal} {\bibinfo  {journal} {AIP Conference Proceedings}\
  }\textbf {\bibinfo {volume} {988}},\ \bibinfo {pages} {128--131} (\bibinfo
  {year} {2008})},\ \Eprint
  {http://arxiv.org/abs/https://aip.scitation.org/doi/pdf/10.1063/1.2905053}
  {https://aip.scitation.org/doi/pdf/10.1063/1.2905053} \BibitemShut {NoStop}%
\bibitem [{\citenamefont {Jardin}(2010)}]{jardin2010methods}%
  \BibitemOpen
  \bibfield  {author} {\bibinfo {author} {\bibfnamefont {S.}~\bibnamefont
  {Jardin}},\ }\href {\doibase 10.1201/EBK1439810958} {\emph {\bibinfo {title}
  {Computational {Methods} in {Plasma} {Physics}}}},\ Vol.\ \bibinfo {volume}
  {1st ed.}\ (\bibinfo  {publisher} {CRC Press},\ \bibinfo {year} {2010})\
  \Eprint {http://arxiv.org/abs/https://doi.org/10.1201/EBK1439810958}
  {https://doi.org/10.1201/EBK1439810958} \BibitemShut {NoStop}%
\bibitem [{\citenamefont {Arbon}, \citenamefont {Candy},\ and\ \citenamefont
  {Belli}(2020)}]{arbon2020mxh}%
  \BibitemOpen
  \bibfield  {author} {\bibinfo {author} {\bibfnamefont {R.}~\bibnamefont
  {Arbon}}, \bibinfo {author} {\bibfnamefont {J.}~\bibnamefont {Candy}}, \ and\
  \bibinfo {author} {\bibfnamefont {E.~A.}\ \bibnamefont {Belli}},\ }\bibfield
  {title} {\enquote {\bibinfo {title} {Rapidly-convergent flux-surface shape
  parameterization},}\ }\href@noop {} {\bibfield  {journal} {\bibinfo
  {journal} {Plasma Physics and Controlled Fusion}\ }\textbf {\bibinfo {volume}
  {63}},\ \bibinfo {pages} {012001} (\bibinfo {year} {2020})}\BibitemShut
  {NoStop}%
\bibitem [{\citenamefont {McClenaghan}\ \emph {et~al.}(2024)\citenamefont
  {McClenaghan}, \citenamefont {Akçay}, \citenamefont {Amara}, \citenamefont
  {Sun}, \citenamefont {Madireddy}, \citenamefont {Lao}, \citenamefont
  {Kruger},\ and\ \citenamefont {Meneghini}}]{mcclenaghan2024efitai}%
  \BibitemOpen
  \bibfield  {author} {\bibinfo {author} {\bibfnamefont {J.}~\bibnamefont
  {McClenaghan}}, \bibinfo {author} {\bibfnamefont {C.}~\bibnamefont {Akçay}},
  \bibinfo {author} {\bibfnamefont {T.~B.}\ \bibnamefont {Amara}}, \bibinfo
  {author} {\bibfnamefont {X.}~\bibnamefont {Sun}}, \bibinfo {author}
  {\bibfnamefont {S.}~\bibnamefont {Madireddy}}, \bibinfo {author}
  {\bibfnamefont {L.~L.}\ \bibnamefont {Lao}}, \bibinfo {author} {\bibfnamefont
  {S.~E.}\ \bibnamefont {Kruger}}, \ and\ \bibinfo {author} {\bibfnamefont
  {O.~M.}\ \bibnamefont {Meneghini}},\ }\bibfield  {title} {\enquote {\bibinfo
  {title} {Augmenting machine learning of {Grad–Shafranov} equilibrium
  reconstruction with {Green's} functions},}\ }\href {\doibase
  10.1063/5.0213625} {\bibfield  {journal} {\bibinfo  {journal} {Physics of
  Plasmas}\ }\textbf {\bibinfo {volume} {31}},\ \bibinfo {pages} {082507}
  (\bibinfo {year} {2024})},\ \Eprint
  {http://arxiv.org/abs/https://pubs.aip.org/aip/pop/article-pdf/doi/10.1063/5.0213625/20106149/082507\_1\_5.0213625.pdf}
  {https://pubs.aip.org/aip/pop/article-pdf/doi/10.1063/5.0213625/20106149/082507\_1\_5.0213625.pdf}
  \BibitemShut {NoStop}%
\bibitem [{\citenamefont {Meneghini}\ \emph {et~al.}(2024)\citenamefont
  {Meneghini}, \citenamefont {Slendebroek}, \citenamefont {Lyons},
  \citenamefont {McLaughlin}, \citenamefont {McClenaghan}, \citenamefont
  {Stagner}, \citenamefont {Harvey}, \citenamefont {Neiser}, \citenamefont
  {Ghiozzi}, \citenamefont {Dose}, \citenamefont {Guterl}, \citenamefont
  {Zalzali}, \citenamefont {Cote}, \citenamefont {Shi}, \citenamefont
  {Weisberg}, \citenamefont {Smith}, \citenamefont {Grierson},\ and\
  \citenamefont {Candy}}]{meneghini2024fuse}%
  \BibitemOpen
  \bibfield  {author} {\bibinfo {author} {\bibfnamefont {O.}~\bibnamefont
  {Meneghini}}, \bibinfo {author} {\bibfnamefont {T.}~\bibnamefont
  {Slendebroek}}, \bibinfo {author} {\bibfnamefont {B.~C.}\ \bibnamefont
  {Lyons}}, \bibinfo {author} {\bibfnamefont {K.}~\bibnamefont {McLaughlin}},
  \bibinfo {author} {\bibfnamefont {J.}~\bibnamefont {McClenaghan}}, \bibinfo
  {author} {\bibfnamefont {L.}~\bibnamefont {Stagner}}, \bibinfo {author}
  {\bibfnamefont {J.}~\bibnamefont {Harvey}}, \bibinfo {author} {\bibfnamefont
  {T.~F.}\ \bibnamefont {Neiser}}, \bibinfo {author} {\bibfnamefont
  {A.}~\bibnamefont {Ghiozzi}}, \bibinfo {author} {\bibfnamefont
  {G.}~\bibnamefont {Dose}}, \bibinfo {author} {\bibfnamefont {J.}~\bibnamefont
  {Guterl}}, \bibinfo {author} {\bibfnamefont {A.}~\bibnamefont {Zalzali}},
  \bibinfo {author} {\bibfnamefont {T.}~\bibnamefont {Cote}}, \bibinfo {author}
  {\bibfnamefont {N.}~\bibnamefont {Shi}}, \bibinfo {author} {\bibfnamefont
  {D.}~\bibnamefont {Weisberg}}, \bibinfo {author} {\bibfnamefont {S.~P.}\
  \bibnamefont {Smith}}, \bibinfo {author} {\bibfnamefont {B.~A.}\ \bibnamefont
  {Grierson}}, \ and\ \bibinfo {author} {\bibfnamefont {J.}~\bibnamefont
  {Candy}},\ }\href {https://arxiv.org/abs/2409.05894} {\enquote {\bibinfo
  {title} {{FUSE} (fusion synthesis engine): {A} {Next} {Generation}
  {Framework} for {Integrated} {Design} of {Fusion Pilot Plants}},}\ }
  (\bibinfo {year} {2024}),\ \Eprint {http://arxiv.org/abs/2409.05894}
  {arXiv:2409.05894 [physics.plasm-ph]} \BibitemShut {NoStop}%
\bibitem [{\citenamefont {Van~Zeeland}\ \emph {et~al.}(2017)\citenamefont
  {Van~Zeeland}, \citenamefont {Carlstrom}, \citenamefont {Finkenthal},
  \citenamefont {Boivin}, \citenamefont {Colio}, \citenamefont {Du},
  \citenamefont {Gattuso}, \citenamefont {Glass}, \citenamefont {Muscatello},
  \citenamefont {O’Neill} \emph {et~al.}}]{van2017tests}%
  \BibitemOpen
  \bibfield  {author} {\bibinfo {author} {\bibfnamefont {M.}~\bibnamefont
  {Van~Zeeland}}, \bibinfo {author} {\bibfnamefont {T.}~\bibnamefont
  {Carlstrom}}, \bibinfo {author} {\bibfnamefont {D.}~\bibnamefont
  {Finkenthal}}, \bibinfo {author} {\bibfnamefont {R.}~\bibnamefont {Boivin}},
  \bibinfo {author} {\bibfnamefont {A.}~\bibnamefont {Colio}}, \bibinfo
  {author} {\bibfnamefont {D.}~\bibnamefont {Du}}, \bibinfo {author}
  {\bibfnamefont {A.}~\bibnamefont {Gattuso}}, \bibinfo {author} {\bibfnamefont
  {F.}~\bibnamefont {Glass}}, \bibinfo {author} {\bibfnamefont
  {C.}~\bibnamefont {Muscatello}}, \bibinfo {author} {\bibfnamefont
  {R.}~\bibnamefont {O’Neill}},  \emph {et~al.},\ }\bibfield  {title}
  {\enquote {\bibinfo {title} {Tests of a two-color interferometer and
  polarimeter for {ITER} density measurements},}\ }\href@noop {} {\bibfield
  {journal} {\bibinfo  {journal} {Plasma Physics and Controlled Fusion}\
  }\textbf {\bibinfo {volume} {59}},\ \bibinfo {pages} {125005} (\bibinfo
  {year} {2017})}\BibitemShut {NoStop}%
\bibitem [{\citenamefont {Nelson}, \citenamefont {Ford},\ and\ \citenamefont
  {Payne}(2013)}]{nelson2013run}%
  \BibitemOpen
  \bibfield  {author} {\bibinfo {author} {\bibfnamefont {B.}~\bibnamefont
  {Nelson}}, \bibinfo {author} {\bibfnamefont {E.~B.}\ \bibnamefont {Ford}}, \
  and\ \bibinfo {author} {\bibfnamefont {M.~J.}\ \bibnamefont {Payne}},\
  }\bibfield  {title} {\enquote {\bibinfo {title} {Run dmc: an efficient,
  parallel code for analyzing radial velocity observations using n-body
  integrations and differential evolution markov chain monte carlo},}\
  }\href@noop {} {\bibfield  {journal} {\bibinfo  {journal} {The Astrophysical
  Journal Supplement Series}\ }\textbf {\bibinfo {volume} {210}},\ \bibinfo
  {pages} {11} (\bibinfo {year} {2013})}\BibitemShut {NoStop}%
\bibitem [{\citenamefont {Ter~Braak}\ and\ \citenamefont
  {Vrugt}(2008)}]{ter2008differential}%
  \BibitemOpen
  \bibfield  {author} {\bibinfo {author} {\bibfnamefont {C.~J.}\ \bibnamefont
  {Ter~Braak}}\ and\ \bibinfo {author} {\bibfnamefont {J.~A.}\ \bibnamefont
  {Vrugt}},\ }\bibfield  {title} {\enquote {\bibinfo {title} {Differential
  evolution markov chain with snooker updater and fewer chains},}\ }\href@noop
  {} {\bibfield  {journal} {\bibinfo  {journal} {Statistics and Computing}\
  }\textbf {\bibinfo {volume} {18}},\ \bibinfo {pages} {435--446} (\bibinfo
  {year} {2008})}\BibitemShut {NoStop}%
\bibitem [{\citenamefont {Foreman-Mackey}(2016)}]{corner}%
  \BibitemOpen
  \bibfield  {author} {\bibinfo {author} {\bibfnamefont {D.}~\bibnamefont
  {Foreman-Mackey}},\ }\bibfield  {title} {\enquote {\bibinfo {title}
  {corner.py: Scatterplot matrices in python},}\ }\href {\doibase
  10.21105/joss.00024} {\bibfield  {journal} {\bibinfo  {journal} {The Journal
  of Open Source Software}\ }\textbf {\bibinfo {volume} {1}},\ \bibinfo {pages}
  {24} (\bibinfo {year} {2016})}\BibitemShut {NoStop}%
\bibitem [{\citenamefont {Uzun-Kaymak}\ \emph {et~al.}(2024)\citenamefont
  {Uzun-Kaymak}, \citenamefont {Galante}, \citenamefont {Foley},\ and\
  \citenamefont {Levinton}}]{uzun2024designing}%
  \BibitemOpen
  \bibfield  {author} {\bibinfo {author} {\bibfnamefont {I.}~\bibnamefont
  {Uzun-Kaymak}}, \bibinfo {author} {\bibfnamefont {M.}~\bibnamefont
  {Galante}}, \bibinfo {author} {\bibfnamefont {E.}~\bibnamefont {Foley}}, \
  and\ \bibinfo {author} {\bibfnamefont {F.}~\bibnamefont {Levinton}},\
  }\bibfield  {title} {\enquote {\bibinfo {title} {Designing {ITER} motional
  stark effect line shift (mse-ls) spectrometers},}\ }\href@noop {} {\bibfield
  {journal} {\bibinfo  {journal} {Review of Scientific Instruments}\ }\textbf
  {\bibinfo {volume} {95}} (\bibinfo {year} {2024})}\BibitemShut {NoStop}%
\bibitem [{\citenamefont {Imazawa}, \citenamefont {Ono},\ and\ \citenamefont
  {Hatae}(2023)}]{PoPola}%
  \BibitemOpen
  \bibfield  {author} {\bibinfo {author} {\bibfnamefont {R.}~\bibnamefont
  {Imazawa}}, \bibinfo {author} {\bibfnamefont {T.}~\bibnamefont {Ono}}, \ and\
  \bibinfo {author} {\bibfnamefont {T.}~\bibnamefont {Hatae}},\ }\bibfield
  {title} {\enquote {\bibinfo {title} {Design of optical transmission line of
  {ITER} poloidal polarimeter},}\ }\href@noop {} {\bibfield  {journal}
  {\bibinfo  {journal} {Fusion Engineering and Design}\ }\textbf {\bibinfo
  {volume} {192}},\ \bibinfo {pages} {113607} (\bibinfo {year}
  {2023})}\BibitemShut {NoStop}%
\end{thebibliography}%
\end{document}